\documentclass[aps,prl,twocolumn,longbibliography]{revtex4-2}
\usepackage[pdftex]{graphicx}
\usepackage[colorlinks=true,citecolor=blue]{hyperref}
\usepackage{graphicx,textcomp,dcolumn,hyperref,upgreek}
\usepackage[normalem]{ulem}

\usepackage{physics}

\DeclareMathAlphabet\mathbfcal{OMS}{cmsy}{b}{n}

\usepackage{url}

\usepackage{dsfont}

\usepackage{blindtext}

\usepackage{float}

\newcommand{\en}{\varepsilon}

\newcommand{\stiff}{A}

\newcommand\varpm{\mathbin{\vcenter{\hbox{%
  \oalign{\hfil$\scriptstyle-$\hfil\cr
          \noalign{\kern-.3ex}
          $\scriptscriptstyle({+})$\cr}%
}}}}

\begin{document}

\title{
Entangling color centers via magnon-antimagnon pair creation
}
\author{Eric~Kleinherbers}
\email{ekleinherbers@physics.ucla.edu}
\affiliation{Department of Physics and Astronomy and Bhaumik Institute for Theoretical Physics, University of California, Los Angeles, California 90095, USA}

\author{Shane~P.~Kelly}
\affiliation{Department of Physics and Astronomy and Bhaumik Institute for Theoretical Physics, University of California, Los Angeles, California 90095, USA}

\author{Yaroslav~Tserkovnyak}
\affiliation{Department of Physics and Astronomy and Bhaumik Institute for Theoretical Physics, University of California, Los Angeles, California 90095, USA}
               
\date{\today}

\begin{abstract} 
We present how entanglement between a spatially separated pair of color centers can be created by letting them weakly interact with the quantum fluctuations of a nonequilibrium magnetic environment. 
To this end, we consider two coupled ferromagnets, one in the ground state and one in an inverted state with respect to an applied magnetic field. The resulting energetic instability leads to a quantum spin current in the vacuum state that is sustained by the creation of magnon-antimagnon pairs at the interface. We show that these quantum fluctuations imprint a steady-state entanglement onto the two dipole-coupled color centers through nonlocal dissipation. We derive conditions for establishing a maximally entangled Bell state. This entanglement is absent in thermal equilibrium. 
\end{abstract}

\maketitle

\textit{Introduction.}---Engineering entanglement between qubits is a common goal in all of quantum science as it is key for quantum advantage in sensing, computation, and communication.
Embedding qubits into a solid-state environment~\cite{chatterjee_2021,trifunovic_2012,doherty_2013,nagy_2019,gottscholl_2020,burkard_2023} has potential to advance this endeavor due to the plethora of environments available to engineer nonlocal correlations through electronic and spintronic control. 
Here, our goal is to imprint these correlations onto weakly coupled qubits in the form of entanglement through natural dissipative evolution~\cite{zou_2022}. For qubits, we consider color centers (e.g., nitrogen-vacancy centers), which are used as pristine tools for magnetometry~\cite{casola_2018} and relaxometry~\cite{mzyk_2022} even at room temperature. 

\begin{figure}[t]
\includegraphics[width=.45\textwidth]{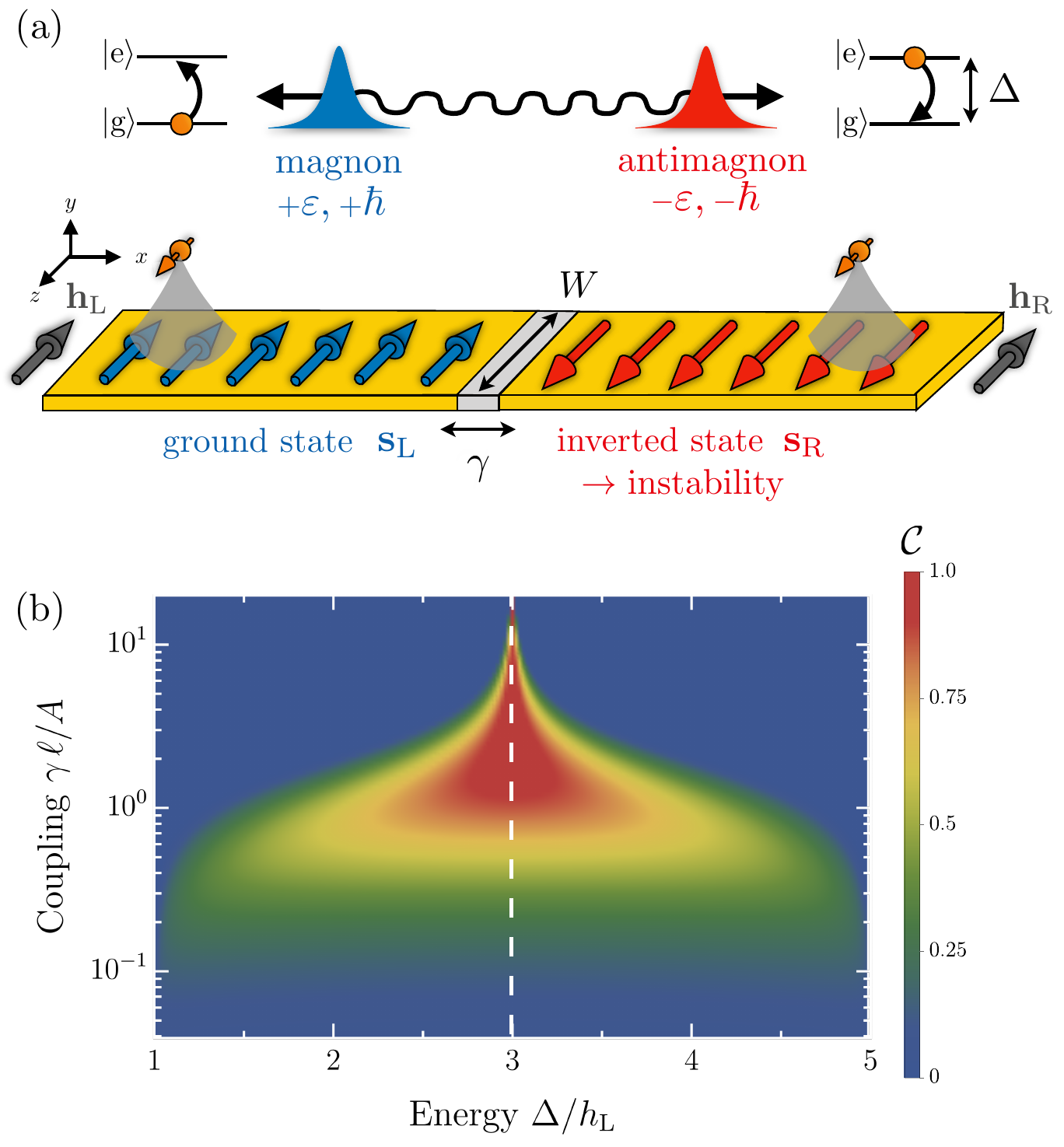}
	\caption{(a) Setup for magnon-antimagnon pair creation at the interface between two ferromagnets of width $W$, one in the ground state (left) and one in the inverted state (right).  The ferromagnets are magnetically coupled to color centers (orange arrows) that can absorb the generated magnon-antimagnon pairs and become entangled with each other. (b) Steady-state concurrence ${\cal C}$ (an entanglement measure) as a function of the color center energy $\Delta$ and the interfacial coupling $\gamma$. We choose $h_\text{R}=5 h_\text{L}$ and define the length scale $\ell=\sqrt{\stiff/(h_\text{R}{-}h_\text{L})}$. 
 }
	\label{fig:1}
\end{figure}
A necessary condition for creating \textit{steady-state} entanglement between independent qubits that are weakly coupled through an environment is to drive the environment out of equilibrium.
This is because in equilibrium --- irrespective of how strong the environment correlations are --- detailed balance will always enforce an uncorrelated Gibbs state~\cite{schaller_2011}. Only transient entanglement generation may occur~\cite{zou_2022}.
Therefore, we consider in this Letter the \textit{nonequilibrium}  magnetic environment shown in Fig.~\ref{fig:1}(a), which was studied in Ref.~\cite{harms_2022} as a magnonic version of the bosonic Klein paradox~\cite{mayoral_2011,wagner_2010}. A ground-state ferromagnet (left) is weakly coupled at the interface to an \textit{inverted-state} ferromagnet (right), where the former is aligned and the latter is antialigned with the applied magnetic field. To avoid the instability, the inverted system is dynamically stabilized by a spin torque~\cite{harms_2022}. 

Conceptually, the physics is similar to an analog black hole~\cite{balbinot_2013,nunez_2017}, where the horizon is mimicked by the interface between the stable and the unstable magnet with positive and negative excitation energies, respectively. Due to quantum fluctuations, this instability can be harvested through magnon pair production that is reminiscent of Hawking radiation. Here, the pair creation sustains a purely quantum spin current consisting of positive-energy magnons  with $+\hbar$ emitted to the left and negative-energy magnons (or antimagnons~\cite{harms_2024}) with $-\hbar$ emitted to the right. 

Our main result is demonstrating that two color centers, placed in proximity of the two magnets, pick up the entanglement of the magnon-antimagnon pairs in the nonequilibrium steady state. The anisotropic magnetostatic coupling between the dipole moments of the color centers and the stray field of the magnets turns out to be beneficial, as it facilitates a perfectly chiral coupling and eliminates dephasing. We derive a Lindblad master equation with nonlocal gain and loss and find that in the steady state, the color centers become entangled, see Fig.~\ref{fig:1}(b). The concurrence ${\cal C}$~\cite{wooters_1998}, used as a measure of entanglement, is close to one in a sweet spot (red) tuned by the color center energy $\Delta$ and the weak interfacial exchange coupling strength $\gamma$. Along the symmetry axis (white dashed line), the steady state becomes pure and is of the simple form
\begin{align}
\ket{\psi}=\frac{r \ket{\text{g,e}}-t \ket{\text{e,g}}}{\sqrt{{r}^2+{t}^2}},\label{eq:steadystate}
\end{align}
where $g$ and $e$ denote the ground and the excited state of the two color centers. 
Here, ${r}$ and ${t}$ are scattering amplitudes describing the pair creation in the environment  (see details below). They are obtained from a generalized (nonunitary) scattering theory~\cite{balbinot_2013} using a bosonic Bogoliubov ansatz for both positive-energy and negative-energy magnons. The scattering amplitudes fulfill ${r}^2-{t}^2=1$. 
For zero interfacial coupling, $\gamma=0$, the production of magnon-antimagnon pairs is zero, which leads to ${t}=0$ and ${r}=1$. In this limit,  we obtain the unentangled steady state $\ket{\psi}=\ket{\text{g,e}}$, where the spontaneous emission of a negative-energy magnon triggers a population inversion on the right side. As the coupling is turned on, $\gamma>0$, entangled magnon pairs are produced, ${t}>0$,  and  can nonlocally excite the color centers through dissipative coupling.  In the limit ${r}\approx {t}\gg1$, the steady state approaches the Bell state $\ket{\psi}\approx\left( \ket{\text{g,e}}-\ket{\text{e,g}}\right)/{\sqrt{2}}$.

\begin{figure*}[t]
\includegraphics[width=.9\textwidth]{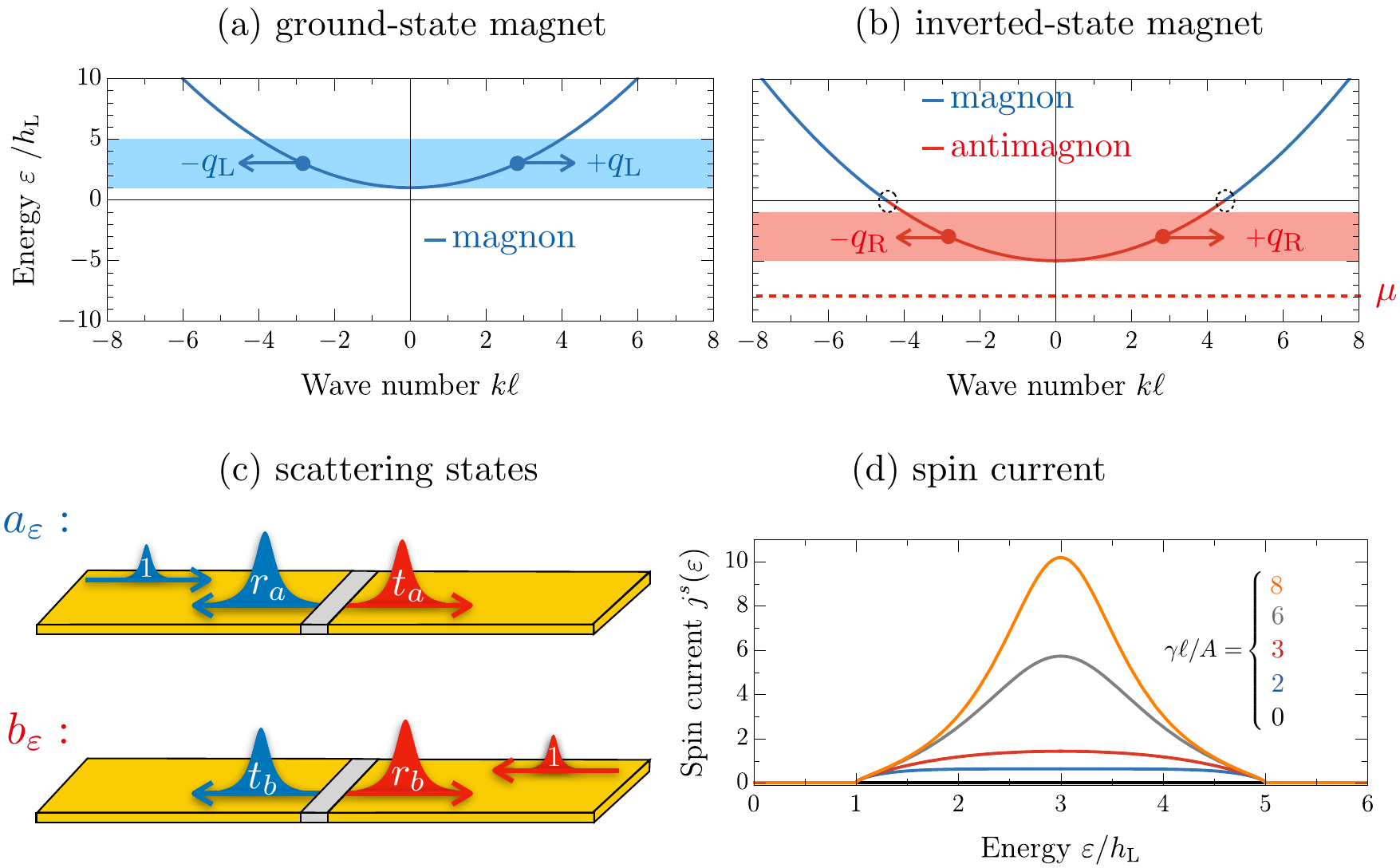}
	\caption{Magnon excitation energies $\varepsilon$ for (a) the left ground-state magnet and (b) the right inverted-state magnet. For energies $h_\text{L}<\varepsilon<h_\text{R}$ (blue area on the left)  and $-h_\text{R}<{\varepsilon}<-h_\text{L}$ (red area on the right) magnons with positive and negative energy can be created pairwise at the interface. For stabilization, the right magnet is pumped with spin accumulation $\mu<-h_\text{R}$ (red dashed line). The dashed circles indicate the zero-energy magnons that potentially can induce dephasing. (c) Sketch of the scattering state $a_\varepsilon$  (${b}_{\varepsilon}$) constructed from an incoming wave with amplitude $1$ and outgoing  waves with reflection amplitude $r_a$ ($r_b$)  and  transmission amplitude $t_a$ ($t_b$). (d) The magnon pair creation sustains a spin current $j^s(\varepsilon)$ in the vacuum state $\ket{\Omega}$, which is flowing to the left. The current is only nonzero for energies $h_\text{L}<\varepsilon<h_\text{R}$. The stronger the coupling $\gamma$ at the interface, the more pairs are emitted. The parameters are the same as in Fig.~\ref{fig:1}.  }
	\label{fig:2}
\end{figure*}

 \textit{Nonequilibrium magnetic environment.}---We model the nonequilibrium magnetic environment via the Hamiltonian  $H_\text{E}=\int\mathrm{d}x\left({\cal H }_\text{L}+{\cal H }_\text{R}\right)+{H }_\text{I}$  that consists of contributions from the left ground-state ferromagnet, the right inverted-state ferromagnet, and the coupling at the interface (see Fig.~\ref{fig:1}). In the absence of ${ H }_\text{I}$, the left (right) magnet is in the ground (inverted) state, $\ev{\vb{s}}_{\text{L}(\text{R})}=\varpm s\vb{e}_z$, aligned (antialigned) with the external field $\vb{h}_{\text{L}(\text{R})}=-{h}_{\text{L}(\text{R})}\vb{e}_z/\hbar $, where $s$ is the linear spin density.
Spin excitations are captured by the magnonic field operators $\psi_\text{L}= \left(s_\text{L}^x-is^y_\text{L}\right)/\sqrt{2\hbar s}$ and $\psi_\text{R}=\left(s_\text{R}^x+is^y_\text{R}\right)/\sqrt{2\hbar s}$ with spin $+\hbar$ and $-\hbar$ on top of the ground ($x<0$) and the inverted state ($x>0$), respectively.  
Then, the bulk Hamiltonian densities become 
\begin{align}
{\cal H }_\alpha=A \left(\partial_x \psi^\dagger_\alpha\right) \left(\partial_x \psi_\alpha\right)  \pm  {h}_\alpha  \psi^\dagger_\alpha \psi_\alpha,\label{eq:bulkham}
\end{align}
where  $\alpha\in\text{L,R}$. The first term reflects the bulk exchange interaction with spin stiffness $\stiff$ that favors a parallel alignment of the spins.  The second term is the Zeeman energy with locally applied external fields $h_\text{L}$ and $h_\text{R}$.  We remark that as a practical realization, the fields $h_\text{L}$ and $h_\text{R}$ can alternatively be created through the exchange bias~\cite{berkowitz_1999} or by an easy- and a hard-axis anisotropy, respectively.  
The resulting bulk excitation energies of the magnons are given by $\varepsilon=\stiff k^2 \pm h_\text{L,R}$ [see Fig.~\ref{fig:2}(a)-(b)] as a function of the wavenumber $k$.  The magnon energies are raised by $h_\text{L}$ in the left magnet, while they are lowered by $h_\text{R}$ in the right magnet. As a result, we obtain in the inverted magnet negative excitation energies, $\varepsilon<0$, which indicates the instability. In fact, dissipation through Gilbert damping with coefficient $\alpha$ will eventually relax the magnet to its ground state. A dynamical stabilization is achieved, for example, by pumping the inverted magnet with a spin torque $\tau_s$. If the effective spin accumulation $\mu=\hbar \tau_s /\alpha$ lies below the magnon band, $\mu<- h_\text{R}$  [see red dashed line in Fig.~\ref{fig:2}(b)], the system is stable against Gilbert damping~\cite{SM}.

Generation of entangled magnon-antimagnon pairs occurs when the inverted magnet --- which acts as a source of energy --- is coupled at the interface to the normal magnet.
We assume an exchange coupling $\sim \vb{s}_\text{L} \cdot \vb{s}_\text{R}$ that couples the left and right magnons via
\begin{align}
{H }_\text{I}=\gamma \left(\psi_\text{L}^\dagger \psi_\text{L}+{\psi}_\text{R}^\dagger {\psi}_\text{R}+{\psi}_\text{R} \psi_\text{L} +\psi_\text{L}^\dagger {\psi}_\text{R}^\dagger\right)\Big\vert_{x=0},\label{eq:interaction}
\end{align}
where $\gamma$ is the exchange coupling strength. In a linear treatment of the spin fluctuations, the sign of  $\gamma$ is irrelevant. Also, we are interested in the regime of a weak link, where the microscopic exchange energy is much weaker at the interface than in the bulk. Crucially, ${ H }_\text{I}$ contains the pair creation term $\sim  \psi_\text{L}^\dagger \psi_\text{R}^\dagger$ that becomes resonant when positive-energy magnons in the left magnet [blue dots in Fig.~\ref{fig:2}(a)] are matched with negative-energy magnons in the right magnet  [red dots in Fig.~\ref{fig:2}(b)]. This can only happen for energies in the interval $h_\text{L}<\varepsilon<h_\text{R}$ in the left magnet (blue shaded area) and $-h_\text{R}<\varepsilon<-h_\text{L}$ in the right magnet (red shaded area). Thus, we require $h_\text{R}-h_\text{L}>0$.
The last two terms of Eq.~\eqref{eq:interaction} generate two-mode squeezing similar to photon pair creation in parametric amplifiers~\cite{caves_1985,schumaker_1985}.

 \textit{Magnon-antimagnon pair creation.}---Production of entangled magnon-antimagnon pairs is described by a generalized (nonunitary) scattering theory in a bosonic Bogoliubov space. 
The calculations are performed in analogy to Ref.~\cite{balbinot_2013}, where Hawking radiation of phonons in acoustic black holes is investigated. 
Expanding the magnon field operators in terms of inscattering states, we obtain~\cite{SM}
\begin{align}
\psi_\text{L}(x)&= \int \limits_{h_\text{L}}^{h_\text{R}}\frac{\mathrm{d}\varepsilon}{\sqrt{h v_\text{L}}} \Bigg[\left(e^{i q_\text{L}x }{+}r_a e^{{-}i q_\text{L}x }\right) a_\varepsilon +{t}_b e^{{-}i q_\text{L}x }b_{-\varepsilon}^\dagger\Bigg], \label{eq:bogansatz1}\\ 
{\psi}_\text{R}(x)&= \int \limits_{-h_\text{R}}^{-h_\text{L}}\frac{\mathrm{d}\varepsilon}{\sqrt{h {v}_\text{R}}} \Bigg[\left(e^{-i {q}_\text{R}x }{+}{r_b} e^{i {q}_\text{R}x }\right) b_\varepsilon{+}t_a e^{i {q}_\text{R}x }{a}_{{-}\varepsilon}^\dagger\Bigg], \label{eq:bogansatz2}
\end{align}
where $h=2\pi\hbar$ is the Planck constant. The operators $a_\varepsilon$ and $b_\varepsilon$ describe the two relevant inscattering solutions with  $[a_\varepsilon,a^\dagger_{\varepsilon^\prime}]=[b_\varepsilon,b^\dagger_{\varepsilon^\prime}]= \delta(\varepsilon{-}\varepsilon^\prime)$ [see Fig.~\ref{fig:2}(a)-(c)]. 
The scattering mode $a_\varepsilon$  describes an incident positive-energy magnon from the left with amplitude one that gets reflected with amplitude $r_a(\varepsilon)$ and  transmitted as a negative-energy magnon with amplitude $t_a(\varepsilon)$ [see Fig.~\ref{fig:2}(c)].  Scattering into positive-energy magnons on the right is not allowed by spin conservation. 
Similarly, the scattering mode ${b}_\varepsilon$ describes  an incident negative-energy magnon from the right with  amplitude one that gets reflected with amplitude $r_b(\varepsilon)$ and transmitted as a positive-energy magnon with amplitude ${t}_b(\varepsilon)$  [see Fig.~\ref{fig:2}(c)]. 
The wave numbers are given by $q_\text{L}=\sqrt{(\varepsilon{-}h_\text{L})/A}$ and $q_\text{R}=\sqrt{(\varepsilon{+}h_\text{R})/A}$ with group velocities $v_\text{L}= 2Aq_\text{L}/\hbar$ and ${v}_\text{R}= 2A q_\text{R}/\hbar$, respectively. 
We remark that the expansion from Eqs.~\eqref{eq:bogansatz1}-\eqref{eq:bogansatz2} is incomplete, as higher-energy scattering solutions are omitted. These scattering states are irrelevant to our findings below but are fully accounted for in the Supplemental Material~\cite{SM}, where we also present the scattering amplitudes, which fulfill $\abs{{t}_a(-\varepsilon)}=\abs{{t}_b(\varepsilon)}\equiv t$,  $\abs{{r}_a(\varepsilon)}=\abs{{r}_b(-\varepsilon)}\equiv r$, as well as ${r}^2-{t}^2=1$. The last relation lies at the heart of the bosonic Klein paradox~\cite{mayoral_2011,wagner_2010},  where the nonunitarity of the scattering process even leads to ${r}\ge 1$~[indicated in Fig.~\ref{fig:2}(c)]. In contrast, for conventional scattering problems, unitarity enforces ${r}^2+{t}^2=1$.

Having solved for the magnon operators, we can determine the spin current due to the spontaneous pair production via
\begin{align}
    {{I}}^{s}_\alpha(x)= \pm 2\stiff \left[{\psi}_\alpha^\dagger \left(-i \partial_x {\psi}_\alpha\right)-\left(-i\partial_x \psi_\alpha^\dagger\right) {\psi}_\alpha\right],
\end{align}
where the $+$ and the $-$ sign holds for $\alpha=\text{L}$ and $\alpha=\text{R}$, respectively. 
Even in the vacuum state defined through $a_\varepsilon \ket{\Omega}={b}_{\varepsilon} \ket{\Omega}=0$, 
the system sustains a constant spin current that is flowing to the left:
\begin{align}\label{eq:spincurrent}
\ev{ {{I}}^{s}_{\text{L}}(x)}{\Omega}=\ev{ {{I}}^{s}_\text{R}(x)}{\Omega}=-\frac{1}{\pi} \int \limits_{h_\text{L}}^{h_\text{R}}{\mathrm{d}\en} \, t(\en)^2,
\end{align}
which is equal in the left and right magnet and independent of $x$. 
Here, the prefactor follows from $2\hbar/h$, where each magnon carries the spin $\hbar$, in analogy to the electron current in the Landauer-Büttiker formalism~\cite{lesovik_2011}.  
The spin current is purely quantum as it originates in the vacuum fluctuations at the interface. It is expected to be dominant for small temperatures  $ k_\text{B}T\ll h_\text{L}$ and $k_\text{B}T\ll -h_\text{R}-\mu$, where a thermal background can be neglected~\footnote{In a thermal state, where magnons are incoming from $x\rightarrow \pm \infty$, the current from  Eq.~\eqref{eq:spincurrent} would acquire a factor $1+n(\en)+n(-\en-\mu)$ in the integrand with Bose-Einstein distribution $n(\en)=\left(e^{\en/k_\text{B}T} {-}1\right)^{-1}$. Thermal contributions remain small if $ k_\text{B}T\ll h_\text{L}$ and $k_\text{B}T\ll -h_\text{R}-\mu$.}. The vacuum spin current corresponds to the constant creation of magnon-antimagnon pairs, where $+\hbar$ magnons with positive energy  $\varepsilon\in(h_\text{L},h_\text{R})$ move to the left and  $-\hbar$ magnons with negative energy $\varepsilon\in(-h_\text{R},-h_\text{L})$ (or antimagnons) move to the right.   
In Fig.~\ref{fig:2}(d), we show the energy-resolved spin current $j^s(\varepsilon)= \, t ^2/\pi$ of positive-energy magnons moving to the left as a measure of the spontaneous pair emission spectrum. The emission occurs symmetrically with respect to $\varepsilon=(h_\text{L}{+}h_\text{R})/2$.
A significant change of the spin current $j^s$ occurs as the exchange coupling $\gamma$ is varied on the characteristic scale ${A/\ell}$ with $\ell=\sqrt{A/(h_\text{R}{-}h_\text{L})}$. 
In a microscopic Heisenberg model, this characteristic value $\gamma \sim {A/\ell}$ translates to $J^\prime \ell \sim J a$, where $J$ and $J^\prime$ are the bulk and interfacial exchange energies and $a$ is the atomic lattice constant. Since $\ell\gg a$, a weak link with $J^\prime\ll J$ is sufficient to enable spontaneous pair emission with $j^s\sim {\cal O}(1)$. 
In a recent work~\cite{bassant_2024}, the current-current correlations were analyzed to investigate the entanglement between magnons and antimagnons.

\textit{Imprinting entanglement.}---To sense the pair creation, we couple two color centers to the magnet. Their bare Hamiltonian is given by
$H_\text{S} = - \Delta\sigma^z_{\text{L}}/2  -  \Delta\sigma^z_{\text{R}}/2$, with $\Delta>0$, using the Pauli matrices $\sigma_\alpha^{j}$ labeled by $j\in x,y,z$. Thus, energetically, both color centers favor the $+z$ direction, with ground state $\ket{\text{g,g}}$. The basis states are denoted by $\ket{\text{g,g}} ,\ket{\text{g,e}} ,\ket{\text{e,g}}$ and $\ket{\text{e,e}}$ with $\text{g}$ and $\text{e}$ standing for the ground and the excited state, respectively. 

The color centers interact with the stray field $\vb{h}_s$ of the environment through the magnetostatic coupling Hamiltonian
\begin{align}
    H_\text{C}=- \frac{\hbar}{2} \sum\limits_{\alpha=\text{L,R}}  \Bigg[ h^+_s(\vb{r}_\alpha)  \sigma^-_\alpha&+h^-_s(\vb{r}_\alpha)  \sigma^+_\alpha \Bigg],  \label{eq:coupling}
\end{align}
where we introduced $\sigma_\alpha^\pm=\left( \sigma^x_\alpha \pm i \sigma^y_\alpha\right)/2$ and $h_s^\pm= h^x_s\pm i h^y_s$.
Here, $\vb{r}_\alpha=(x_\alpha,d,0)$ denotes the position of the color centers at height $d>0$ on either side of the interface with $x_\text{L}<0$ and  $x_\text{R}>0$ (see Fig.~\ref{fig:1}). The stray field $\vb{h}_s$ can be calculated via the magnetostatic Greens function~\cite{guslienko_2011}, where its source are the magnonic excitations described by $\psi_\text{L}$ and ${\psi}_\text{R}$ of Eqs.~\eqref{eq:bogansatz1}-\eqref{eq:bogansatz2}.

\begin{figure}[t]
\includegraphics[width=.45\textwidth]{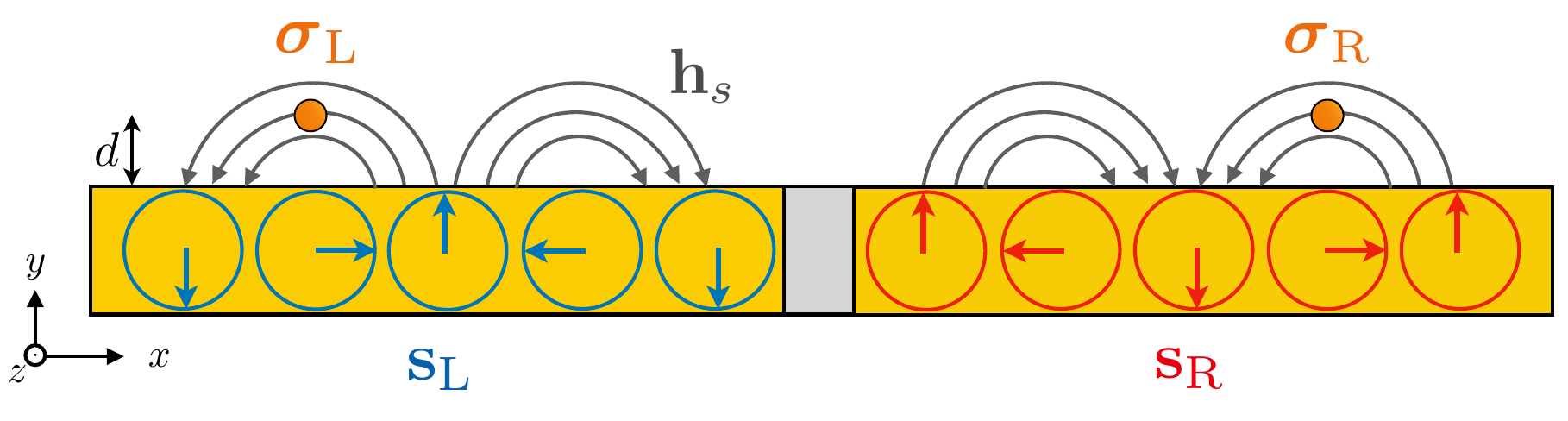}
	\caption{Snapshot of Halbach-like stray field $\vb{h}_s$ from the magnetic excitations described by $\vb{s}_\text{L}$ and $\vb{s}_\text{R}$. The stray magnetic field is only created above the magnet if magnons are moving to the left (right) on the left (right) side. In addition, a combination of time reversal symmetry and mirror symmetry in the $xy$ plane ensures $h_s^z=0$. To sense the pair creation, the two color centers (orange) are placed above the magnet at height $d$,  one on each side of the interface (gray) with $x_\text{L}<0$ and $x_\text{R}>0$.}
	\label{fig:3}
\end{figure}

Generically, the anisotropy of the magnetostatic coupling from Eq.~\eqref{eq:coupling} is expected to be a nuisance, as it spoils the spin $z$ symmetry and can lead to magnon-induced dephasing [which is resonant at zero energy and therefore energetically enabled by the instability in the right magnet, see dashed circles in Fig.~\ref{fig:2}(b)~\footnote{If we included anisotropy in the magnet, we would obtain anomalous terms~\cite{erickson_1991} resulting in elliptical polarization of the magnons and the emergence of exceptional points at zero energy~\cite{harms_2024}. We expect the dephasing of the color centers, which is resonant at zero energy, to be sensitive to this physics.  }].
However, in our setup, the coupling turns out to be beneficial for two reasons.
First, due to a combination of time-reversal and $xy$ mirror symmetry of the magnonic spin excitations (see Fig.~\ref{fig:3}), it follows that $h_s^z=0$. This eliminates magnon-induced dephasing of the color centers. 
Second, as shown in Fig.~\ref{fig:3}, the stray field of a single magnon is Halbach array-like~\cite{halbach_1985}, so it is nonzero either only above or below the magnet depending on its wave vector $k$. In order to be sensitive to the pair creation, we want the left color center to couple to left-moving positive-energy magnons ($k=-q_\text{L}$) and the right color center to couple to right-moving negative-energy magnons ($k=+{q}_\text{R}$). For $\abs{x_\alpha}\gg d$, we obtain~\cite{SM}
\begin{align}
h^+_s(\vb{r}_\text{L})&\propto \int \limits_{h_\text{L}}^{h_\text{R}}\frac{\mathrm{d}\varepsilon}{\sqrt{h  v_\text{L}} } q_\text{L} e^{-q_\text{L} d}  e^{-  i q_\text{L} x_\text{L}}  \Bigg(r_a a_\varepsilon + {t}_b {b_{-\varepsilon}^\dagger}\Bigg), \\
h^-_s(\vb{r}_\text{R})&\propto\!\!\int \limits_{-h_\text{R}}^{-h_\text{L}}\! \frac{\mathrm{d}\varepsilon}{\sqrt{h{v}_\text{R}} } {q}_\text{R} e^{-{q}_\text{R} d}  e^{i {q}_\text{R} x_\text{R}}  \Bigg(  r_b b_\varepsilon+{t}_a a^\dagger_{-\varepsilon}\Bigg).  \label{eq:strayfield}
\end{align}
Thus, coupling to the stray field introduces the typical filter function  $q_\text{L} e^{-q_\text{L} d}$ and  ${q}_\text{R} e^{-{q}_\text{R}d}$~\cite{flebus_2018,guslienko_2011}, which makes the color centers insensitive to long-wavelength magnons~\footnote{This justifies neglecting magnetic dipole interactions in the bulk magnet of Eq.~\eqref{eq:bulkham}, as they give rise to long-wavelength modifications~\cite{karenowska_2014}.}. The chirality has eliminated the incoming plane waves with amplitude one from Eqs.~\eqref{eq:bogansatz1}-\eqref{eq:bogansatz2}.

Integrating out the environment degrees of freedom $a_\varepsilon$ and ${b}_\varepsilon$ in the effective vacuum state $\ket{\Omega}$, we obtain a Lindblad master equation for the reduced density matrix $\rho$ in a Born-Markov treatment~\cite{SM}
\begin{align}
\partial_t \rho = \frac{1}{i\hbar} \left[ H_\text{S}+\delta H,\rho\right] + \sum_{i=1}^2 \Gamma_0 \left( L_i \rho L_i^\dagger -\frac{1}{2} \{ L^\dagger_i L_i, \rho \} \right),\label{eq:lindblad}
\end{align}
where the rate is given by  $\Gamma_0=\gamma_e^4 \hbar^2 s/W^2 A d$ in Gaussian units with the gyromagnetic ratio $\gamma_e$ and the width of the magnetic strip $W$.
The magnetic fluctuations encoded in Greens functions of $h^+_s(\vb{r}_\alpha)$ and  $h^-_s(\vb{r}_\alpha)$ generically induce both nonlocal coherent dynamics described by $\delta H$ as well as nonlocal dissipative dynamics described by the Lindblad operators $L_1$ and $L_2$. For the chiral magnetostatic coupling, it can be shown that  $\delta H=0$~\footnote{There is, however, a local Lamb shift  $\propto \delta_\alpha \sigma_\alpha^z$ on either side, which renormalizes the bare energies $\Delta$. The dynamics will be sensitive to differences $\delta_\text{L}-\delta_\text{R}$.}.  
For the nonlocal dissipation, we obtain the Lindblad operators
\begin{align}
 L_1&=   \sqrt{q_\text{L}d}\, e^{-q_\text{L} d} r\, {\sigma}^+_\text{L}+ \sqrt{\bar{q}_\text{R}d}\,e^{-\bar{q}_\text{R} d}  t \,{\sigma}^+_\text{R}, \\
  L_2&=  \sqrt{q_\text{L}d}\, e^{-q_\text{L} d} t\,{\sigma}^-_\text{L}+ \sqrt{\bar{q}_\text{R}d}\,e^{-\bar{q}_\text{R} d}  r \,{\sigma}^-_\text{R},
\end{align}
where all functions are evaluated at the resonant energy $\varepsilon=\Delta$ and we define $\bar{q}_\text{R}(\Delta)={q}_\text{R}(-\Delta)$. In addition, we performed local $z$ rotations to obtain the real amplitudes $r$ and $t$. The dependence on the positions of the color centers $x_\alpha$ has disappeared due to the chiral nature of the coupling. 
 
The Lindblad operator $L_1$ ($L_2$) describes the emission (absorption) of a positive-energy magnon on the left side paired with the absorption (emission) of a negative-energy magnon on the right side, with both processes decreasing (increasing) the energy of the color centers. 
In the limit of no pair creation, $\gamma\rightarrow0$, we obtain $r=1$ and $t=0$. Then, $L_1$ and $L_2$ trigger the spontaneous emission of a positive-energy magnon on the left side and a negative-energy on the right side. Thus, the steady state of the color centers becomes $\ket{g,e}$, where on the right side, a population inversion is created akin to negative temperature. 
For finite coupling, $\gamma>0$,  magnon-antimagnon pairs are produced, which enables both absorption with $t>0$ and stimulated emission with $r>1$. Then, the environment can imprint entanglement by admixing the $\ket{e,g}$ state. In Fig.~\ref{fig:1}(b), we show the concurrence ${\cal C}$ of the steady state that follows from Eq.~\eqref{eq:lindblad}. For zero coupling, $\gamma=0$, we get ${\cal C}=0$, as the steady state is $\ket{g,e}$. However, as the coupling $\gamma$ is turned on, entanglement is created and we observe a sweet spot [red in Fig.~\ref{fig:1}(b)] with a concurrence close to one, where the interfacial coupling is at around $\gamma \sim \sqrt{A (h_\text{R}{-}h_\text{L})}$ and $\Delta\sim (h_\text{L}{+}h_\text{R})/2$.   As $\gamma$ increases beyond the sweet spot, the pair creation appears more classical and entanglement only occurs in a narrow window around $\Delta=(h_\text{L}{+}h_\text{R})/2$. At this tuned position we obtain $q_\text{L}=\bar{q}_\text{R}$ and the steady state becomes pure and has a particularly simple form given by Eq.~\eqref{eq:steadystate}. The concurrence in this case reduces to ${\cal C}=2rt/({r}^2{+}{t}^2)$. 
We remark that the chiral nature of the coupling is not essential to create entanglement. Without chirality, however, the spatial dependence of the scattering solution survives~\cite{SM}.

\textit{Conclusion.}---In this Letter, we demonstrated that steady-state entanglement can be imprinted onto a pair of color centers by letting them weakly interact with a magnetic solid-state environment. 
A necessary ingredient is to drive the environment out of equilibrium, which is achieved by an inverted magnet that can host negative-energy magnons (or antimagnons). At the interface to a normal magnet, positive-energy and negative-energy magnons are created in entangled pairs. Via a chiral magnetostatic coupling to the stray field of the magnet, the two color centers can pick up the magnon-antimagnon pairs through nonlocal absorption and emission processes. In a sweet spot, tuned by the color center energy and the interfacial exchange coupling, the color centers naturally evolve into an entangled steady state that even approaches the Bell state.

A key to imprinting entanglement was the creation of two-mode squeezing between positive-energy and negative-energy magnons. We emphasize that rather than an external drive, the interfacial exchange coupling served as an outlet that can harvest the energy supplied by the bulk inverted magnet. This is contrasted with the parametric drive used to create squeezing in quantum optics~\cite{caves_81, caves_1985,schumaker_1985}.

For future work, it might be interesting to embrace the full nonlinearity of the spin dynamics in the environment in order to prepare non-Gaussian states, going beyond two-mode squeezing.
Furthermore, it seems promising to couple different magnetic systems to the inverted region that can be used as a source of energy to induce interesting entangled dynamics.  
Yet another promising route is to scale up the effect by considering two-dimensional generalizations, 
which allows an extended ensemble of color centers placed at interfaces between ground- and inverted-state regions.

\begin{acknowledgments}
The authors thank Jamir Marino and Rembert Duine for fruitful discussions. This work was supported by the U.S. Department of Energy, Office of Basic Energy Sciences under Grant No. DE-SC0012190.
\end{acknowledgments}

\bibliography{References}

\begin{thebibliography}{33}%
\makeatletter
\providecommand \@ifxundefined [1]{%
 \@ifx{#1\undefined}
}%
\providecommand \@ifnum [1]{%
 \ifnum #1\expandafter \@firstoftwo
 \else \expandafter \@secondoftwo
 \fi
}%
\providecommand \@ifx [1]{%
 \ifx #1\expandafter \@firstoftwo
 \else \expandafter \@secondoftwo
 \fi
}%
\providecommand \natexlab [1]{#1}%
\providecommand \enquote  [1]{``#1''}%
\providecommand \bibnamefont  [1]{#1}%
\providecommand \bibfnamefont [1]{#1}%
\providecommand \citenamefont [1]{#1}%
\providecommand \href@noop [0]{\@secondoftwo}%
\providecommand \href [0]{\begingroup \@sanitize@url \@href}%
\providecommand \@href[1]{\@@startlink{#1}\@@href}%
\providecommand \@@href[1]{\endgroup#1\@@endlink}%
\providecommand \@sanitize@url [0]{\catcode `\\12\catcode `\$12\catcode
  `\&12\catcode `\#12\catcode `\^12\catcode `\_12\catcode `\%12\relax}%
\providecommand \@@startlink[1]{}%
\providecommand \@@endlink[0]{}%
\providecommand \url  [0]{\begingroup\@sanitize@url \@url }%
\providecommand \@url [1]{\endgroup\@href {#1}{\urlprefix }}%
\providecommand \urlprefix  [0]{URL }%
\providecommand \Eprint [0]{\href }%
\providecommand \doibase [0]{https://doi.org/}%
\providecommand \selectlanguage [0]{\@gobble}%
\providecommand \bibinfo  [0]{\@secondoftwo}%
\providecommand \bibfield  [0]{\@secondoftwo}%
\providecommand \translation [1]{[#1]}%
\providecommand \BibitemOpen [0]{}%
\providecommand \bibitemStop [0]{}%
\providecommand \bibitemNoStop [0]{.\EOS\space}%
\providecommand \EOS [0]{\spacefactor3000\relax}%
\providecommand \BibitemShut  [1]{\csname bibitem#1\endcsname}%
\let\auto@bib@innerbib\@empty
\bibitem [{\citenamefont {Chatterjee}\ \emph {et~al.}(2021)\citenamefont
  {Chatterjee}, \citenamefont {Stevenson}, \citenamefont {De~Franceschi},
  \citenamefont {Morello}, \citenamefont {de~Leon},\ and\ \citenamefont
  {Kuemmeth}}]{chatterjee_2021}%
  \BibitemOpen
  \bibfield  {author} {\bibinfo {author} {\bibfnamefont {A.}~\bibnamefont
  {Chatterjee}}, \bibinfo {author} {\bibfnamefont {P.}~\bibnamefont
  {Stevenson}}, \bibinfo {author} {\bibfnamefont {S.}~\bibnamefont
  {De~Franceschi}}, \bibinfo {author} {\bibfnamefont {A.}~\bibnamefont
  {Morello}}, \bibinfo {author} {\bibfnamefont {N.~P.}\ \bibnamefont
  {de~Leon}},\ and\ \bibinfo {author} {\bibfnamefont {F.}~\bibnamefont
  {Kuemmeth}},\ }\bibfield  {title} {\bibinfo {title} {Semiconductor qubits in
  practice},\ }\href {https://doi.org/10.1038/s42254-021-00283-9} {\bibfield
  {journal} {\bibinfo  {journal} {Nat. Rev. Phys.}\ }\textbf {\bibinfo {volume}
  {3}},\ \bibinfo {pages} {157} (\bibinfo {year} {2021})}\BibitemShut {NoStop}%
\bibitem [{\citenamefont {Trifunovic}\ \emph {et~al.}(2012)\citenamefont
  {Trifunovic}, \citenamefont {Dial}, \citenamefont {Trif}, \citenamefont
  {Wootton}, \citenamefont {Abebe}, \citenamefont {Yacoby},\ and\ \citenamefont
  {Loss}}]{trifunovic_2012}%
  \BibitemOpen
  \bibfield  {author} {\bibinfo {author} {\bibfnamefont {L.}~\bibnamefont
  {Trifunovic}}, \bibinfo {author} {\bibfnamefont {O.}~\bibnamefont {Dial}},
  \bibinfo {author} {\bibfnamefont {M.}~\bibnamefont {Trif}}, \bibinfo {author}
  {\bibfnamefont {J.~R.}\ \bibnamefont {Wootton}}, \bibinfo {author}
  {\bibfnamefont {R.}~\bibnamefont {Abebe}}, \bibinfo {author} {\bibfnamefont
  {A.}~\bibnamefont {Yacoby}},\ and\ \bibinfo {author} {\bibfnamefont
  {D.}~\bibnamefont {Loss}},\ }\bibfield  {title} {\bibinfo {title}
  {Long-distance spin-spin coupling via floating gates},\ }\href
  {https://doi.org/10.1103/PhysRevX.2.011006} {\bibfield  {journal} {\bibinfo
  {journal} {Phys. Rev. X}\ }\textbf {\bibinfo {volume} {2}},\ \bibinfo {pages}
  {011006} (\bibinfo {year} {2012})}\BibitemShut {NoStop}%
\bibitem [{\citenamefont {Doherty}\ \emph {et~al.}(2013)\citenamefont
  {Doherty}, \citenamefont {Manson}, \citenamefont {Delaney}, \citenamefont
  {Jelezko}, \citenamefont {Wrachtrup},\ and\ \citenamefont
  {Hollenberg}}]{doherty_2013}%
  \BibitemOpen
  \bibfield  {author} {\bibinfo {author} {\bibfnamefont {M.~W.}\ \bibnamefont
  {Doherty}}, \bibinfo {author} {\bibfnamefont {N.~B.}\ \bibnamefont {Manson}},
  \bibinfo {author} {\bibfnamefont {P.}~\bibnamefont {Delaney}}, \bibinfo
  {author} {\bibfnamefont {F.}~\bibnamefont {Jelezko}}, \bibinfo {author}
  {\bibfnamefont {J.}~\bibnamefont {Wrachtrup}},\ and\ \bibinfo {author}
  {\bibfnamefont {L.~C.}\ \bibnamefont {Hollenberg}},\ }\bibfield  {title}
  {\bibinfo {title} {The nitrogen-vacancy colour centre in diamond},\ }\href
  {https://doi.org/https://doi.org/10.1016/j.physrep.2013.02.001} {\bibfield
  {journal} {\bibinfo  {journal} {Phys. Rep.}\ }\textbf {\bibinfo {volume}
  {528}},\ \bibinfo {pages} {1} (\bibinfo {year} {2013})}\BibitemShut {NoStop}%
\bibitem [{\citenamefont {Nagy}\ \emph {et~al.}(2019)\citenamefont {Nagy},
  \citenamefont {Niethammer}, \citenamefont {Widmann}, \citenamefont {Chen},
  \citenamefont {Udvarhelyi}, \citenamefont {Bonato}, \citenamefont {Hassan},
  \citenamefont {Karhu}, \citenamefont {Ivanov}, \citenamefont {Son},
  \citenamefont {Maze}, \citenamefont {Ohshima}, \citenamefont {Soykal},
  \citenamefont {Gali}, \citenamefont {Lee}, \citenamefont {Kaiser},\ and\
  \citenamefont {Wrachtrup}}]{nagy_2019}%
  \BibitemOpen
  \bibfield  {author} {\bibinfo {author} {\bibfnamefont {R.}~\bibnamefont
  {Nagy}}, \bibinfo {author} {\bibfnamefont {M.}~\bibnamefont {Niethammer}},
  \bibinfo {author} {\bibfnamefont {M.}~\bibnamefont {Widmann}}, \bibinfo
  {author} {\bibfnamefont {Y.-C.}\ \bibnamefont {Chen}}, \bibinfo {author}
  {\bibfnamefont {P.}~\bibnamefont {Udvarhelyi}}, \bibinfo {author}
  {\bibfnamefont {C.}~\bibnamefont {Bonato}}, \bibinfo {author} {\bibfnamefont
  {J.~U.}\ \bibnamefont {Hassan}}, \bibinfo {author} {\bibfnamefont
  {R.}~\bibnamefont {Karhu}}, \bibinfo {author} {\bibfnamefont {I.~G.}\
  \bibnamefont {Ivanov}}, \bibinfo {author} {\bibfnamefont {N.~T.}\
  \bibnamefont {Son}}, \bibinfo {author} {\bibfnamefont {J.~R.}\ \bibnamefont
  {Maze}}, \bibinfo {author} {\bibfnamefont {T.}~\bibnamefont {Ohshima}},
  \bibinfo {author} {\bibfnamefont {{\"O}.~O.}\ \bibnamefont {Soykal}},
  \bibinfo {author} {\bibfnamefont {{\'A}.}~\bibnamefont {Gali}}, \bibinfo
  {author} {\bibfnamefont {S.-Y.}\ \bibnamefont {Lee}}, \bibinfo {author}
  {\bibfnamefont {F.}~\bibnamefont {Kaiser}},\ and\ \bibinfo {author}
  {\bibfnamefont {J.}~\bibnamefont {Wrachtrup}},\ }\bibfield  {title} {\bibinfo
  {title} {High-fidelity spin and optical control of single silicon-vacancy
  centres in silicon carbide},\ }\href
  {https://doi.org/10.1038/s41467-019-09873-9} {\bibfield  {journal} {\bibinfo
  {journal} {Nat. Commun.}\ }\textbf {\bibinfo {volume} {10}},\ \bibinfo
  {pages} {1954} (\bibinfo {year} {2019})}\BibitemShut {NoStop}%
\bibitem [{\citenamefont {Gottscholl}\ \emph {et~al.}(2020)\citenamefont
  {Gottscholl}, \citenamefont {Kianinia}, \citenamefont {Soltamov},
  \citenamefont {Orlinskii}, \citenamefont {Mamin}, \citenamefont {Bradac},
  \citenamefont {Kasper}, \citenamefont {Krambrock}, \citenamefont {Sperlich},
  \citenamefont {Toth}, \citenamefont {Aharonovich},\ and\ \citenamefont
  {Dyakonov}}]{gottscholl_2020}%
  \BibitemOpen
  \bibfield  {author} {\bibinfo {author} {\bibfnamefont {A.}~\bibnamefont
  {Gottscholl}}, \bibinfo {author} {\bibfnamefont {M.}~\bibnamefont
  {Kianinia}}, \bibinfo {author} {\bibfnamefont {V.}~\bibnamefont {Soltamov}},
  \bibinfo {author} {\bibfnamefont {S.}~\bibnamefont {Orlinskii}}, \bibinfo
  {author} {\bibfnamefont {G.}~\bibnamefont {Mamin}}, \bibinfo {author}
  {\bibfnamefont {C.}~\bibnamefont {Bradac}}, \bibinfo {author} {\bibfnamefont
  {C.}~\bibnamefont {Kasper}}, \bibinfo {author} {\bibfnamefont
  {K.}~\bibnamefont {Krambrock}}, \bibinfo {author} {\bibfnamefont
  {A.}~\bibnamefont {Sperlich}}, \bibinfo {author} {\bibfnamefont
  {M.}~\bibnamefont {Toth}}, \bibinfo {author} {\bibfnamefont {I.}~\bibnamefont
  {Aharonovich}},\ and\ \bibinfo {author} {\bibfnamefont {V.}~\bibnamefont
  {Dyakonov}},\ }\bibfield  {title} {\bibinfo {title} {Initialization and
  read-out of intrinsic spin defects in a van der {W}aals crystal at room
  temperature},\ }\href {https://doi.org/10.1038/s41563-020-0619-6} {\bibfield
  {journal} {\bibinfo  {journal} {Nat. Mater.}\ }\textbf {\bibinfo {volume}
  {19}},\ \bibinfo {pages} {540} (\bibinfo {year} {2020})}\BibitemShut
  {NoStop}%
\bibitem [{\citenamefont {Burkard}\ \emph {et~al.}(2023)\citenamefont
  {Burkard}, \citenamefont {Ladd}, \citenamefont {Pan}, \citenamefont
  {Nichol},\ and\ \citenamefont {Petta}}]{burkard_2023}%
  \BibitemOpen
  \bibfield  {author} {\bibinfo {author} {\bibfnamefont {G.}~\bibnamefont
  {Burkard}}, \bibinfo {author} {\bibfnamefont {T.~D.}\ \bibnamefont {Ladd}},
  \bibinfo {author} {\bibfnamefont {A.}~\bibnamefont {Pan}}, \bibinfo {author}
  {\bibfnamefont {J.~M.}\ \bibnamefont {Nichol}},\ and\ \bibinfo {author}
  {\bibfnamefont {J.~R.}\ \bibnamefont {Petta}},\ }\bibfield  {title} {\bibinfo
  {title} {Semiconductor spin qubits},\ }\href
  {https://doi.org/10.1103/RevModPhys.95.025003} {\bibfield  {journal}
  {\bibinfo  {journal} {Rev. Mod. Phys.}\ }\textbf {\bibinfo {volume} {95}},\
  \bibinfo {pages} {025003} (\bibinfo {year} {2023})}\BibitemShut {NoStop}%
\bibitem [{\citenamefont {Zou}\ \emph {et~al.}(2022)\citenamefont {Zou},
  \citenamefont {Zhang},\ and\ \citenamefont {Tserkovnyak}}]{zou_2022}%
  \BibitemOpen
  \bibfield  {author} {\bibinfo {author} {\bibfnamefont {J.}~\bibnamefont
  {Zou}}, \bibinfo {author} {\bibfnamefont {S.}~\bibnamefont {Zhang}},\ and\
  \bibinfo {author} {\bibfnamefont {Y.}~\bibnamefont {Tserkovnyak}},\
  }\bibfield  {title} {\bibinfo {title} {Bell-state generation for spin qubits
  via dissipative coupling},\ }\href
  {https://doi.org/10.1103/PhysRevB.106.L180406} {\bibfield  {journal}
  {\bibinfo  {journal} {Phys. Rev. B}\ }\textbf {\bibinfo {volume} {106}},\
  \bibinfo {pages} {L180406} (\bibinfo {year} {2022})}\BibitemShut {NoStop}%
\bibitem [{\citenamefont {Casola}\ \emph {et~al.}(2018)\citenamefont {Casola},
  \citenamefont {van~der Sar},\ and\ \citenamefont {Yacoby}}]{casola_2018}%
  \BibitemOpen
  \bibfield  {author} {\bibinfo {author} {\bibfnamefont {F.}~\bibnamefont
  {Casola}}, \bibinfo {author} {\bibfnamefont {T.}~\bibnamefont {van~der
  Sar}},\ and\ \bibinfo {author} {\bibfnamefont {A.}~\bibnamefont {Yacoby}},\
  }\bibfield  {title} {\bibinfo {title} {Probing condensed matter physics with
  magnetometry based on nitrogen-vacancy centres in diamond},\ }\href
  {https://doi.org/10.1038/natrevmats.2017.88} {\bibfield  {journal} {\bibinfo
  {journal} {Nat. Rev. Mater.}\ }\textbf {\bibinfo {volume} {3}},\ \bibinfo
  {pages} {17088} (\bibinfo {year} {2018})}\BibitemShut {NoStop}%
\bibitem [{\citenamefont {Mzyk}\ \emph {et~al.}(2022)\citenamefont {Mzyk},
  \citenamefont {Sigaeva},\ and\ \citenamefont {Schirhagl}}]{mzyk_2022}%
  \BibitemOpen
  \bibfield  {author} {\bibinfo {author} {\bibfnamefont {A.}~\bibnamefont
  {Mzyk}}, \bibinfo {author} {\bibfnamefont {A.}~\bibnamefont {Sigaeva}},\ and\
  \bibinfo {author} {\bibfnamefont {R.}~\bibnamefont {Schirhagl}},\ }\bibfield
  {title} {\bibinfo {title} {Relaxometry with nitrogen vacancy ({NV}) centers
  in diamond},\ }\href {https://doi.org/10.1021/acs.accounts.2c00520}
  {\bibfield  {journal} {\bibinfo  {journal} {Acc. Chem. Res.}\ }\textbf
  {\bibinfo {volume} {55}},\ \bibinfo {pages} {3572} (\bibinfo {year}
  {2022})}\BibitemShut {NoStop}%
\bibitem [{\citenamefont {Schaller}(2011)}]{schaller_2011}%
  \BibitemOpen
  \bibfield  {author} {\bibinfo {author} {\bibfnamefont {G.}~\bibnamefont
  {Schaller}},\ }\bibfield  {title} {\bibinfo {title} {Quantum equilibration
  under constraints and transport balance},\ }\href
  {https://doi.org/10.1103/PhysRevE.83.031111} {\bibfield  {journal} {\bibinfo
  {journal} {Phys. Rev. E}\ }\textbf {\bibinfo {volume} {83}},\ \bibinfo
  {pages} {031111} (\bibinfo {year} {2011})}\BibitemShut {NoStop}%
\bibitem [{\citenamefont {Harms}\ \emph {et~al.}(2022)\citenamefont {Harms},
  \citenamefont {Yuan},\ and\ \citenamefont {Duine}}]{harms_2022}%
  \BibitemOpen
  \bibfield  {author} {\bibinfo {author} {\bibfnamefont {J.}~\bibnamefont
  {Harms}}, \bibinfo {author} {\bibfnamefont {H.}~\bibnamefont {Yuan}},\ and\
  \bibinfo {author} {\bibfnamefont {R.~A.}\ \bibnamefont {Duine}},\ }\bibfield
  {title} {\bibinfo {title} {Enhanced magnon spin current using the bosonic
  {K}lein paradox},\ }\href {https://doi.org/10.1103/PhysRevApplied.18.064026}
  {\bibfield  {journal} {\bibinfo  {journal} {Phys. Rev. Appl.}\ }\textbf
  {\bibinfo {volume} {18}},\ \bibinfo {pages} {064026} (\bibinfo {year}
  {2022})}\BibitemShut {NoStop}%
\bibitem [{\citenamefont {Mayoral}\ \emph {et~al.}(2011)\citenamefont
  {Mayoral}, \citenamefont {Fabbri},\ and\ \citenamefont
  {Rinaldi}}]{mayoral_2011}%
  \BibitemOpen
  \bibfield  {author} {\bibinfo {author} {\bibfnamefont {C.}~\bibnamefont
  {Mayoral}}, \bibinfo {author} {\bibfnamefont {A.}~\bibnamefont {Fabbri}},\
  and\ \bibinfo {author} {\bibfnamefont {M.}~\bibnamefont {Rinaldi}},\
  }\bibfield  {title} {\bibinfo {title} {{Steplike discontinuities in
  Bose-Einstein condensates and Hawking radiation: Dispersion effects}},\
  }\href {https://doi.org/10.1103/PhysRevD.83.124047} {\bibfield  {journal}
  {\bibinfo  {journal} {Phys. Rev. D}\ }\textbf {\bibinfo {volume} {83}},\
  \bibinfo {pages} {124047} (\bibinfo {year} {2011})}\BibitemShut {NoStop}%
\bibitem [{\citenamefont {Wagner}\ \emph {et~al.}(2010)\citenamefont {Wagner},
  \citenamefont {Ware}, \citenamefont {Su},\ and\ \citenamefont
  {Grobe}}]{wagner_2010}%
  \BibitemOpen
  \bibfield  {author} {\bibinfo {author} {\bibfnamefont {R.~E.}\ \bibnamefont
  {Wagner}}, \bibinfo {author} {\bibfnamefont {M.~R.}\ \bibnamefont {Ware}},
  \bibinfo {author} {\bibfnamefont {Q.}~\bibnamefont {Su}},\ and\ \bibinfo
  {author} {\bibfnamefont {R.}~\bibnamefont {Grobe}},\ }\bibfield  {title}
  {\bibinfo {title} {Bosonic analog of the {K}lein paradox},\ }\href
  {https://doi.org/10.1103/PhysRevA.81.024101} {\bibfield  {journal} {\bibinfo
  {journal} {Phys. Rev. A}\ }\textbf {\bibinfo {volume} {81}},\ \bibinfo
  {pages} {024101} (\bibinfo {year} {2010})}\BibitemShut {NoStop}%
\bibitem [{\citenamefont {Balbinot}\ \emph {et~al.}(2013)\citenamefont
  {Balbinot}, \citenamefont {Carusotto}, \citenamefont {Fabbri}, \citenamefont
  {Mayoral},\ and\ \citenamefont {Recati}}]{balbinot_2013}%
  \BibitemOpen
  \bibfield  {author} {\bibinfo {author} {\bibfnamefont {R.}~\bibnamefont
  {Balbinot}}, \bibinfo {author} {\bibfnamefont {I.}~\bibnamefont {Carusotto}},
  \bibinfo {author} {\bibfnamefont {A.}~\bibnamefont {Fabbri}}, \bibinfo
  {author} {\bibfnamefont {C.}~\bibnamefont {Mayoral}},\ and\ \bibinfo {author}
  {\bibfnamefont {A.}~\bibnamefont {Recati}},\ }\bibinfo {title} {Understanding
  hawking radiation from simple models of atomic bose-einstein condensates},\
  in\ \href {https://doi.org/10.1007/978-3-319-00266-8_9} {\emph {\bibinfo
  {booktitle} {Analogue Gravity Phenomenology: Analogue Spacetimes and
  Horizons, from Theory to Experiment}}},\ \bibinfo {editor} {edited by\
  \bibinfo {editor} {\bibfnamefont {D.}~\bibnamefont {Faccio}}, \bibinfo
  {editor} {\bibfnamefont {F.}~\bibnamefont {Belgiorno}}, \bibinfo {editor}
  {\bibfnamefont {S.}~\bibnamefont {Cacciatori}}, \bibinfo {editor}
  {\bibfnamefont {V.}~\bibnamefont {Gorini}}, \bibinfo {editor} {\bibfnamefont
  {S.}~\bibnamefont {Liberati}},\ and\ \bibinfo {editor} {\bibfnamefont
  {U.}~\bibnamefont {Moschella}}}\ (\bibinfo  {publisher} {Springer
  International Publishing},\ \bibinfo {address} {Cham},\ \bibinfo {year}
  {2013})\ pp.\ \bibinfo {pages} {181--219}\BibitemShut {NoStop}%
\bibitem [{\citenamefont {Rold\'an-Molina}\ \emph {et~al.}(2017)\citenamefont
  {Rold\'an-Molina}, \citenamefont {Nunez},\ and\ \citenamefont
  {Duine}}]{nunez_2017}%
  \BibitemOpen
  \bibfield  {author} {\bibinfo {author} {\bibfnamefont {A.}~\bibnamefont
  {Rold\'an-Molina}}, \bibinfo {author} {\bibfnamefont {A.~S.}\ \bibnamefont
  {Nunez}},\ and\ \bibinfo {author} {\bibfnamefont {R.~A.}\ \bibnamefont
  {Duine}},\ }\bibfield  {title} {\bibinfo {title} {Magnonic black holes},\
  }\href {https://doi.org/10.1103/PhysRevLett.118.061301} {\bibfield  {journal}
  {\bibinfo  {journal} {Phys. Rev. Lett.}\ }\textbf {\bibinfo {volume} {118}},\
  \bibinfo {pages} {061301} (\bibinfo {year} {2017})}\BibitemShut {NoStop}%
\bibitem [{\citenamefont {Harms}\ \emph {et~al.}(2024)\citenamefont {Harms},
  \citenamefont {Yuan},\ and\ \citenamefont {Duine}}]{harms_2024}%
  \BibitemOpen
  \bibfield  {author} {\bibinfo {author} {\bibfnamefont {J.~S.}\ \bibnamefont
  {Harms}}, \bibinfo {author} {\bibfnamefont {H.~Y.}\ \bibnamefont {Yuan}},\
  and\ \bibinfo {author} {\bibfnamefont {R.~A.}\ \bibnamefont {Duine}},\
  }\bibfield  {title} {\bibinfo {title} {{Antimagnonics}},\ }\href
  {https://doi.org/10.1063/5.0151652} {\bibfield  {journal} {\bibinfo
  {journal} {AIP Adv.}\ }\textbf {\bibinfo {volume} {14}},\ \bibinfo {pages}
  {025303} (\bibinfo {year} {2024})}\BibitemShut {NoStop}%
\bibitem [{\citenamefont {Wootters}(1998)}]{wooters_1998}%
  \BibitemOpen
  \bibfield  {author} {\bibinfo {author} {\bibfnamefont {W.~K.}\ \bibnamefont
  {Wootters}},\ }\bibfield  {title} {\bibinfo {title} {Entanglement of
  formation of an arbitrary state of two qubits},\ }\href
  {https://doi.org/10.1103/PhysRevLett.80.2245} {\bibfield  {journal} {\bibinfo
   {journal} {Phys. Rev. Lett.}\ }\textbf {\bibinfo {volume} {80}},\ \bibinfo
  {pages} {2245} (\bibinfo {year} {1998})}\BibitemShut {NoStop}%
\bibitem [{\citenamefont {Berkowitz}\ and\ \citenamefont
  {Takano}(1999)}]{berkowitz_1999}%
  \BibitemOpen
  \bibfield  {author} {\bibinfo {author} {\bibfnamefont {A.}~\bibnamefont
  {Berkowitz}}\ and\ \bibinfo {author} {\bibfnamefont {K.}~\bibnamefont
  {Takano}},\ }\bibfield  {title} {\bibinfo {title} {Exchange anisotropy — a
  review},\ }\href
  {https://doi.org/https://doi.org/10.1016/S0304-8853(99)00453-9} {\bibfield
  {journal} {\bibinfo  {journal} {J. Magn. Magn. Mater.}\ }\textbf {\bibinfo
  {volume} {200}},\ \bibinfo {pages} {552} (\bibinfo {year}
  {1999})}\BibitemShut {NoStop}%
\bibitem [{SM()}]{SM}%
  \BibitemOpen
  \href@noop {} {}\bibinfo {note} {The Supplemental Material contains a
  detailed derivation of (i) the scattering solutions that are used to describe
  the magnon-antimagnon pair creation, (ii) the Lindblad equation for the
  chiral magnetostatic coupling to the color centers, (iii) the Lindblad
  equation for a nonchiral exchange coupling to the color centers.}\BibitemShut
  {Stop}%
\bibitem [{\citenamefont {Caves}\ and\ \citenamefont
  {Schumaker}(1985)}]{caves_1985}%
  \BibitemOpen
  \bibfield  {author} {\bibinfo {author} {\bibfnamefont {C.~M.}\ \bibnamefont
  {Caves}}\ and\ \bibinfo {author} {\bibfnamefont {B.~L.}\ \bibnamefont
  {Schumaker}},\ }\bibfield  {title} {\bibinfo {title} {New formalism for
  two-photon quantum optics. i. quadrature phases and squeezed states},\ }\href
  {https://doi.org/10.1103/PhysRevA.31.3068} {\bibfield  {journal} {\bibinfo
  {journal} {Phys. Rev. A}\ }\textbf {\bibinfo {volume} {31}},\ \bibinfo
  {pages} {3068} (\bibinfo {year} {1985})}\BibitemShut {NoStop}%
\bibitem [{\citenamefont {Schumaker}\ and\ \citenamefont
  {Caves}(1985)}]{schumaker_1985}%
  \BibitemOpen
  \bibfield  {author} {\bibinfo {author} {\bibfnamefont {B.~L.}\ \bibnamefont
  {Schumaker}}\ and\ \bibinfo {author} {\bibfnamefont {C.~M.}\ \bibnamefont
  {Caves}},\ }\bibfield  {title} {\bibinfo {title} {New formalism for
  two-photon quantum optics. ii. mathematical foundation and compact
  notation},\ }\href {https://doi.org/10.1103/PhysRevA.31.3093} {\bibfield
  {journal} {\bibinfo  {journal} {Phys. Rev. A}\ }\textbf {\bibinfo {volume}
  {31}},\ \bibinfo {pages} {3093} (\bibinfo {year} {1985})}\BibitemShut
  {NoStop}%
\bibitem [{\citenamefont {Lesovik}\ and\ \citenamefont
  {Sadovskyy}(2011)}]{lesovik_2011}%
  \BibitemOpen
  \bibfield  {author} {\bibinfo {author} {\bibfnamefont {G.~B.}\ \bibnamefont
  {Lesovik}}\ and\ \bibinfo {author} {\bibfnamefont {I.~A.}\ \bibnamefont
  {Sadovskyy}},\ }\bibfield  {title} {\bibinfo {title} {Scattering matrix
  approach to the description of quantum electron transport},\ }\href
  {https://doi.org/10.3367/UFNe.0181.201110b.1041} {\bibfield  {journal}
  {\bibinfo  {journal} {Phys.-Uspekhi}\ }\textbf {\bibinfo {volume} {54}},\
  \bibinfo {pages} {1007} (\bibinfo {year} {2011})}\BibitemShut {NoStop}%
\bibitem [{Note1()}]{Note1}%
  \BibitemOpen
  \bibinfo {note} {In a thermal state, where magnons are incoming from
  $x\rightarrow \pm \infty $, the current from Eq.~\protect \textup {\hbox
  {\mathsurround \z@ \protect \normalfont (\ignorespaces \ref
  {eq:spincurrent}\unskip \@@italiccorr )}} would acquire a factor
  $1+n(\varepsilon )+n(-\varepsilon -\mu )$ in the integrand with Bose-Einstein
  distribution $n(\varepsilon )=\left (e^{\varepsilon /k_\protect \text {B}T}
  {-}1\right )^{-1}$. Thermal contributions remain small if $ k_\protect \text
  {B}T\ll h_\protect \text {L}$ and $k_\protect \text {B}T\ll -h_\protect \text
  {R}-\mu $.}\BibitemShut {Stop}%
\bibitem [{\citenamefont {Bassant}\ \emph {et~al.}(2024)\citenamefont
  {Bassant}, \citenamefont {Regout}, \citenamefont {Harms},\ and\ \citenamefont
  {Duine}}]{bassant_2024}%
  \BibitemOpen
  \bibfield  {author} {\bibinfo {author} {\bibfnamefont {A.~L.}\ \bibnamefont
  {Bassant}}, \bibinfo {author} {\bibfnamefont {M.~E.~Y.}\ \bibnamefont
  {Regout}}, \bibinfo {author} {\bibfnamefont {J.~S.}\ \bibnamefont {Harms}},\
  and\ \bibinfo {author} {\bibfnamefont {R.~A.}\ \bibnamefont {Duine}},\
  }\bibfield  {title} {\bibinfo {title} {Entangled magnon-pair generation in a
  driven synthetic antiferromagnet},\ }\href
  {https://doi.org/10.1103/PhysRevB.110.094441} {\bibfield  {journal} {\bibinfo
   {journal} {Phys. Rev. B}\ }\textbf {\bibinfo {volume} {110}},\ \bibinfo
  {pages} {094441} (\bibinfo {year} {2024})}\BibitemShut {NoStop}%
\bibitem [{\citenamefont {Guslienko}\ and\ \citenamefont
  {Slavin}(2011)}]{guslienko_2011}%
  \BibitemOpen
  \bibfield  {author} {\bibinfo {author} {\bibfnamefont {K.~Y.}\ \bibnamefont
  {Guslienko}}\ and\ \bibinfo {author} {\bibfnamefont {A.~N.}\ \bibnamefont
  {Slavin}},\ }\bibfield  {title} {\bibinfo {title} {{Magnetostatic Green's
  functions for the description of spin waves in finite rectangular magnetic
  dots and stripes}},\ }\href
  {https://doi.org/https://doi.org/10.1016/j.jmmm.2011.05.020} {\bibfield
  {journal} {\bibinfo  {journal} {J. Magn. Magn. Mater.}\ }\textbf {\bibinfo
  {volume} {323}},\ \bibinfo {pages} {2418} (\bibinfo {year}
  {2011})}\BibitemShut {NoStop}%
\bibitem [{Note2()}]{Note2}%
  \BibitemOpen
  \bibinfo {note} {If we included anisotropy in the magnet, we would obtain
  anomalous terms~\cite {erickson_1991} resulting in elliptical polarization of
  the magnons and the emergence of exceptional points at zero energy~\cite
  {harms_2024}. We expect the dephasing of the color centers, which is resonant
  at zero energy, to be sensitive to this physics.}\BibitemShut {Stop}%
\bibitem [{\citenamefont {Halbach}(1985)}]{halbach_1985}%
  \BibitemOpen
  \bibfield  {author} {\bibinfo {author} {\bibfnamefont {K.}~\bibnamefont
  {Halbach}},\ }\bibfield  {title} {\bibinfo {title} {Application of permanent
  magnets in accelerators and electron storage rings},\ }\href
  {https://doi.org/10.1063/1.335021} {\bibfield  {journal} {\bibinfo  {journal}
  {J. Appl. Phys.}\ }\textbf {\bibinfo {volume} {57}},\ \bibinfo {pages} {3605}
  (\bibinfo {year} {1985})}\BibitemShut {NoStop}%
\bibitem [{\citenamefont {Flebus}\ and\ \citenamefont
  {Tserkovnyak}(2018)}]{flebus_2018}%
  \BibitemOpen
  \bibfield  {author} {\bibinfo {author} {\bibfnamefont {B.}~\bibnamefont
  {Flebus}}\ and\ \bibinfo {author} {\bibfnamefont {Y.}~\bibnamefont
  {Tserkovnyak}},\ }\bibfield  {title} {\bibinfo {title} {Quantum-impurity
  relaxometry of magnetization dynamics},\ }\href
  {https://doi.org/10.1103/PhysRevLett.121.187204} {\bibfield  {journal}
  {\bibinfo  {journal} {Phys. Rev. Lett.}\ }\textbf {\bibinfo {volume} {121}},\
  \bibinfo {pages} {187204} (\bibinfo {year} {2018})}\BibitemShut {NoStop}%
\bibitem [{Note3()}]{Note3}%
  \BibitemOpen
  \bibinfo {note} {This justifies neglecting magnetic dipole interactions in
  the bulk magnet of Eq.~\protect \textup {\hbox {\mathsurround \z@ \protect
  \normalfont (\ignorespaces \ref {eq:bulkham}\unskip \@@italiccorr )}}, as
  they give rise to long-wavelength modifications~\cite
  {karenowska_2014}.}\BibitemShut {Stop}%
\bibitem [{Note4()}]{Note4}%
  \BibitemOpen
  \bibinfo {note} {There is, however, a local Lamb shift $\propto \delta
  _\alpha \sigma _\alpha ^z$ on either side, which renormalizes the bare
  energies $\Delta $. The dynamics will be sensitive to differences $\delta
  _\protect \text {L}-\delta _\protect \text {R}$.}\BibitemShut {Stop}%
\bibitem [{\citenamefont {Caves}(1981)}]{caves_81}%
  \BibitemOpen
  \bibfield  {author} {\bibinfo {author} {\bibfnamefont {C.~M.}\ \bibnamefont
  {Caves}},\ }\bibfield  {title} {\bibinfo {title} {Quantum-mechanical noise in
  an interferometer},\ }\href {https://doi.org/10.1103/PhysRevD.23.1693}
  {\bibfield  {journal} {\bibinfo  {journal} {Phys. Rev. D}\ }\textbf {\bibinfo
  {volume} {23}},\ \bibinfo {pages} {1693} (\bibinfo {year}
  {1981})}\BibitemShut {NoStop}%
\bibitem [{\citenamefont {Erickson}\ and\ \citenamefont
  {Mills}(1991)}]{erickson_1991}%
  \BibitemOpen
  \bibfield  {author} {\bibinfo {author} {\bibfnamefont {R.~P.}\ \bibnamefont
  {Erickson}}\ and\ \bibinfo {author} {\bibfnamefont {D.~L.}\ \bibnamefont
  {Mills}},\ }\bibfield  {title} {\bibinfo {title} {Thermodynamics of thin
  ferromagnetic films in the presence of anisotropy and dipolar coupling},\
  }\href {https://doi.org/10.1103/PhysRevB.44.11825} {\bibfield  {journal}
  {\bibinfo  {journal} {Phys. Rev. B}\ }\textbf {\bibinfo {volume} {44}},\
  \bibinfo {pages} {11825} (\bibinfo {year} {1991})}\BibitemShut {NoStop}%
\bibitem [{\citenamefont {Karenowska}\ \emph {et~al.}(2014)\citenamefont
  {Karenowska}, \citenamefont {Chumak}, \citenamefont {Serga},\ and\
  \citenamefont {Hillebrands}}]{karenowska_2014}%
  \BibitemOpen
  \bibfield  {author} {\bibinfo {author} {\bibfnamefont {A.~D.}\ \bibnamefont
  {Karenowska}}, \bibinfo {author} {\bibfnamefont {A.~V.}\ \bibnamefont
  {Chumak}}, \bibinfo {author} {\bibfnamefont {A.~A.}\ \bibnamefont {Serga}},\
  and\ \bibinfo {author} {\bibfnamefont {B.}~\bibnamefont {Hillebrands}},\
  }\bibinfo {title} {Magnon spintronics},\ in\ \href
  {https://doi.org/10.1007/978-94-007-7604-3_53-1} {\emph {\bibinfo {booktitle}
  {Handbook of Spintronics}}},\ \bibinfo {editor} {edited by\ \bibinfo {editor}
  {\bibfnamefont {Y.}~\bibnamefont {Xu}}, \bibinfo {editor} {\bibfnamefont
  {D.~D.}\ \bibnamefont {Awschalom}},\ and\ \bibinfo {editor} {\bibfnamefont
  {J.}~\bibnamefont {Nitta}}}\ (\bibinfo  {publisher} {Springer Netherlands},\
  \bibinfo {address} {Dordrecht},\ \bibinfo {year} {2014})\ pp.\ \bibinfo
  {pages} {1--38}\BibitemShut {NoStop}%
\end{thebibliography}%


\begin{thebibliography}{7}%
\makeatletter
\providecommand \@ifxundefined [1]{%
 \@ifx{#1\undefined}
}%
\providecommand \@ifnum [1]{%
 \ifnum #1\expandafter \@firstoftwo
 \else \expandafter \@secondoftwo
 \fi
}%
\providecommand \@ifx [1]{%
 \ifx #1\expandafter \@firstoftwo
 \else \expandafter \@secondoftwo
 \fi
}%
\providecommand \natexlab [1]{#1}%
\providecommand \enquote  [1]{``#1''}%
\providecommand \bibnamefont  [1]{#1}%
\providecommand \bibfnamefont [1]{#1}%
\providecommand \citenamefont [1]{#1}%
\providecommand \href@noop [0]{\@secondoftwo}%
\providecommand \href [0]{\begingroup \@sanitize@url \@href}%
\providecommand \@href[1]{\@@startlink{#1}\@@href}%
\providecommand \@@href[1]{\endgroup#1\@@endlink}%
\providecommand \@sanitize@url [0]{\catcode `\\12\catcode `\$12\catcode
  `\&12\catcode `\#12\catcode `\^12\catcode `\_12\catcode `\%12\relax}%
\providecommand \@@startlink[1]{}%
\providecommand \@@endlink[0]{}%
\providecommand \url  [0]{\begingroup\@sanitize@url \@url }%
\providecommand \@url [1]{\endgroup\@href {#1}{\urlprefix }}%
\providecommand \urlprefix  [0]{URL }%
\providecommand \Eprint [0]{\href }%
\providecommand \doibase [0]{https://doi.org/}%
\providecommand \selectlanguage [0]{\@gobble}%
\providecommand \bibinfo  [0]{\@secondoftwo}%
\providecommand \bibfield  [0]{\@secondoftwo}%
\providecommand \translation [1]{[#1]}%
\providecommand \BibitemOpen [0]{}%
\providecommand \bibitemStop [0]{}%
\providecommand \bibitemNoStop [0]{.\EOS\space}%
\providecommand \EOS [0]{\spacefactor3000\relax}%
\providecommand \BibitemShut  [1]{\csname bibitem#1\endcsname}%
\let\auto@bib@innerbib\@empty
\bibitem [{\citenamefont {Harms}\ \emph {et~al.}(2024)\citenamefont {Harms},
  \citenamefont {Yuan},\ and\ \citenamefont {Duine}}]{harms_2024}%
  \BibitemOpen
  \bibfield  {author} {\bibinfo {author} {\bibfnamefont {J.~S.}\ \bibnamefont
  {Harms}}, \bibinfo {author} {\bibfnamefont {H.~Y.}\ \bibnamefont {Yuan}},\
  and\ \bibinfo {author} {\bibfnamefont {R.~A.}\ \bibnamefont {Duine}},\
  }\bibfield  {title} {\bibinfo {title} {{Antimagnonics}},\ }\href
  {https://doi.org/10.1063/5.0151652} {\bibfield  {journal} {\bibinfo
  {journal} {AIP Adv.}\ }\textbf {\bibinfo {volume} {14}},\ \bibinfo {pages}
  {025303} (\bibinfo {year} {2024})}\BibitemShut {NoStop}%
\bibitem [{\citenamefont {Balbinot}\ \emph {et~al.}(2013)\citenamefont
  {Balbinot}, \citenamefont {Carusotto}, \citenamefont {Fabbri}, \citenamefont
  {Mayoral},\ and\ \citenamefont {Recati}}]{balbinot_2013}%
  \BibitemOpen
  \bibfield  {author} {\bibinfo {author} {\bibfnamefont {R.}~\bibnamefont
  {Balbinot}}, \bibinfo {author} {\bibfnamefont {I.}~\bibnamefont {Carusotto}},
  \bibinfo {author} {\bibfnamefont {A.}~\bibnamefont {Fabbri}}, \bibinfo
  {author} {\bibfnamefont {C.}~\bibnamefont {Mayoral}},\ and\ \bibinfo {author}
  {\bibfnamefont {A.}~\bibnamefont {Recati}},\ }\bibinfo {title} {Understanding
  {H}awking radiation from simple models of atomic {B}ose-{E}instein
  condensates},\ in\ \href {https://doi.org/10.1007/978-3-319-00266-8_9} {\emph
  {\bibinfo {booktitle} {Analogue Gravity Phenomenology: Analogue Spacetimes
  and Horizons, from Theory to Experiment}}},\ \bibinfo {editor} {edited by\
  \bibinfo {editor} {\bibfnamefont {D.}~\bibnamefont {Faccio}}, \bibinfo
  {editor} {\bibfnamefont {F.}~\bibnamefont {Belgiorno}}, \bibinfo {editor}
  {\bibfnamefont {S.}~\bibnamefont {Cacciatori}}, \bibinfo {editor}
  {\bibfnamefont {V.}~\bibnamefont {Gorini}}, \bibinfo {editor} {\bibfnamefont
  {S.}~\bibnamefont {Liberati}},\ and\ \bibinfo {editor} {\bibfnamefont
  {U.}~\bibnamefont {Moschella}}}\ (\bibinfo  {publisher} {Springer
  International Publishing},\ \bibinfo {address} {Cham},\ \bibinfo {year}
  {2013})\ pp.\ \bibinfo {pages} {181--219}\BibitemShut {NoStop}%
\bibitem [{\citenamefont {Lesovik}\ and\ \citenamefont
  {Sadovskyy}(2011)}]{lesovik_2011}%
  \BibitemOpen
  \bibfield  {author} {\bibinfo {author} {\bibfnamefont {G.~B.}\ \bibnamefont
  {Lesovik}}\ and\ \bibinfo {author} {\bibfnamefont {I.~A.}\ \bibnamefont
  {Sadovskyy}},\ }\bibfield  {title} {\bibinfo {title} {Scattering matrix
  approach to the description of quantum electron transport},\ }\href
  {https://doi.org/10.3367/UFNe.0181.201110b.1041} {\bibfield  {journal}
  {\bibinfo  {journal} {Phys.-Uspekhi}\ }\textbf {\bibinfo {volume} {54}},\
  \bibinfo {pages} {1007} (\bibinfo {year} {2011})}\BibitemShut {NoStop}%
\bibitem [{\citenamefont {Guslienko}\ and\ \citenamefont
  {Slavin}(2011)}]{guslienko_2011}%
  \BibitemOpen
  \bibfield  {author} {\bibinfo {author} {\bibfnamefont {K.~Y.}\ \bibnamefont
  {Guslienko}}\ and\ \bibinfo {author} {\bibfnamefont {A.~N.}\ \bibnamefont
  {Slavin}},\ }\bibfield  {title} {\bibinfo {title} {{Magnetostatic Green's
  functions for the description of spin waves in finite rectangular magnetic
  dots and stripes}},\ }\href
  {https://doi.org/https://doi.org/10.1016/j.jmmm.2011.05.020} {\bibfield
  {journal} {\bibinfo  {journal} {J. Magn. Magn. Mater.}\ }\textbf {\bibinfo
  {volume} {323}},\ \bibinfo {pages} {2418} (\bibinfo {year}
  {2011})}\BibitemShut {NoStop}%
\bibitem [{\citenamefont {Halbach}(1985)}]{halbach_1985}%
  \BibitemOpen
  \bibfield  {author} {\bibinfo {author} {\bibfnamefont {K.}~\bibnamefont
  {Halbach}},\ }\bibfield  {title} {\bibinfo {title} {Application of permanent
  magnets in accelerators and electron storage rings},\ }\href
  {https://doi.org/10.1063/1.335021} {\bibfield  {journal} {\bibinfo  {journal}
  {J. Appl. Phys.}\ }\textbf {\bibinfo {volume} {57}},\ \bibinfo {pages} {3605}
  (\bibinfo {year} {1985})}\BibitemShut {NoStop}%
\bibitem [{\citenamefont {Zou}\ \emph {et~al.}(2022)\citenamefont {Zou},
  \citenamefont {Zhang},\ and\ \citenamefont {Tserkovnyak}}]{zou_2022}%
  \BibitemOpen
  \bibfield  {author} {\bibinfo {author} {\bibfnamefont {J.}~\bibnamefont
  {Zou}}, \bibinfo {author} {\bibfnamefont {S.}~\bibnamefont {Zhang}},\ and\
  \bibinfo {author} {\bibfnamefont {Y.}~\bibnamefont {Tserkovnyak}},\
  }\bibfield  {title} {\bibinfo {title} {Bell-state generation for spin qubits
  via dissipative coupling},\ }\href
  {https://doi.org/10.1103/PhysRevB.106.L180406} {\bibfield  {journal}
  {\bibinfo  {journal} {Phys. Rev. B}\ }\textbf {\bibinfo {volume} {106}},\
  \bibinfo {pages} {L180406} (\bibinfo {year} {2022})}\BibitemShut {NoStop}%
\bibitem [{\citenamefont {Wootters}(1998)}]{wooters_1998}%
  \BibitemOpen
  \bibfield  {author} {\bibinfo {author} {\bibfnamefont {W.~K.}\ \bibnamefont
  {Wootters}},\ }\bibfield  {title} {\bibinfo {title} {Entanglement of
  formation of an arbitrary state of two qubits},\ }\href
  {https://doi.org/10.1103/PhysRevLett.80.2245} {\bibfield  {journal} {\bibinfo
   {journal} {Phys. Rev. Lett.}\ }\textbf {\bibinfo {volume} {80}},\ \bibinfo
  {pages} {2245} (\bibinfo {year} {1998})}\BibitemShut {NoStop}%
\end{thebibliography}%

\end{document}


\title{
Entangling color centers via magnon-antimagnon pair creation
}
\author{Eric~Kleinherbers}
\email{ekleinherbers@physics.ucla.edu}
\affiliation{Department of Physics and Astronomy and Bhaumik Institute for Theoretical Physics, University of California, Los Angeles, California 90095, USA}

\author{Shane~P.~Kelly}
\affiliation{Department of Physics and Astronomy and Bhaumik Institute for Theoretical Physics, University of California, Los Angeles, California 90095, USA}

\author{Yaroslav~Tserkovnyak}
\affiliation{Department of Physics and Astronomy and Bhaumik Institute for Theoretical Physics, University of California, Los Angeles, California 90095, USA}
               
\date{\today}

\maketitle

\onecolumngrid

{
  \hypersetup{linkcolor=black}
  \vspace{-1cm}
  \tableofcontents
}


\begin{figure*}[t]
\includegraphics[width=13cm]{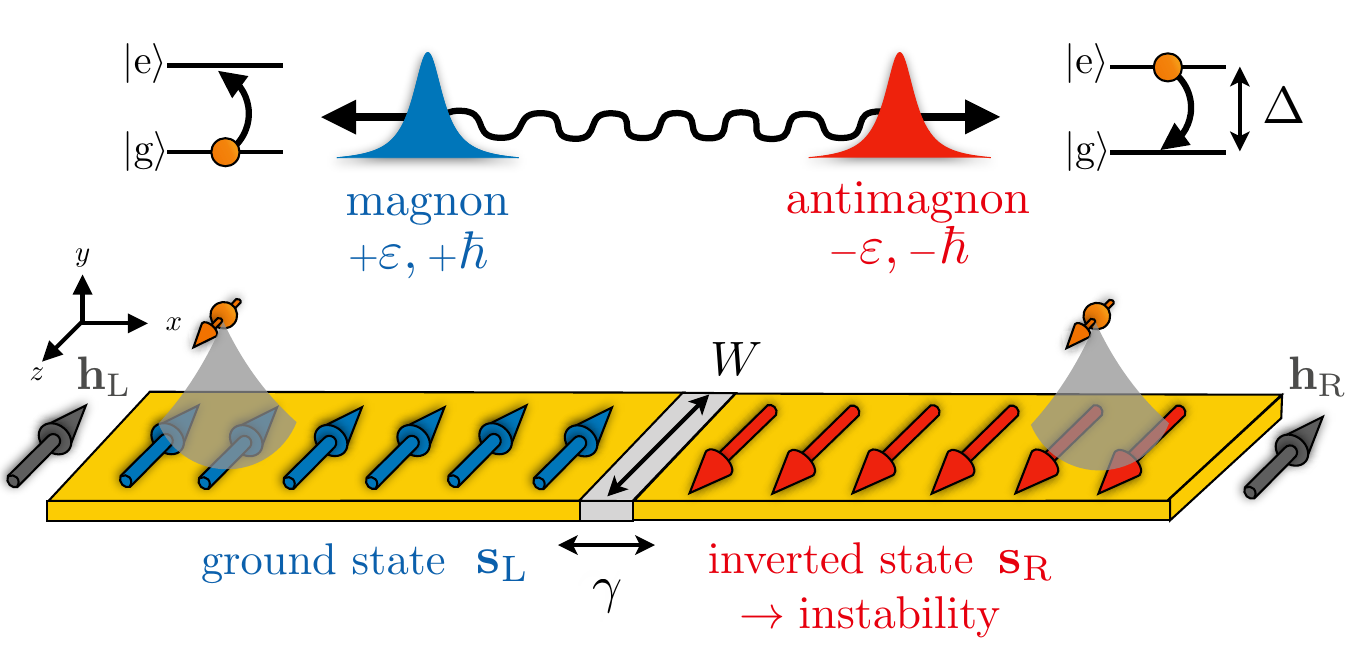}
	\caption{Setup for pair creation of magnons (blue) and antimagnons (red) at an interface between two ferromagnets, one in the ground state (left) and one in an excited state (right). The latter is dynamically stabilized by spin transfer torques.}
	\label{fig:supp1}
\end{figure*}

\section{\label{sec:fluct} Magnon-antimagnon pair creation}
\subsection{\label{sec:not} Notation}
In the Letter, we introduced a more streamlined notation as we ignored unessential scattering solutions. In the Supplemental Material, our description is more complete and therefore uses a slightly different notation. The mapping between notations is given by
\begin{table}[H]
\centering
\begin{tabular}{c|c|c|c|c|c|c}
Letter& $a_\varepsilon$&$b^\dagger_{-\varepsilon}$ &$r_a(\varepsilon)$ &$t_a(-\varepsilon)$ &${r}_b(-\varepsilon) $ &$t_b(\varepsilon)$  \\  \hline
Supplementary  & $a_1(\varepsilon)$& ${a}_2(\varepsilon)$&$ {r}_1(\varepsilon)$&$ \bar{t}_1(\varepsilon) $&$ \bar{r}_2(\varepsilon) $&${t}_2(\varepsilon)$  \\\end{tabular}
\end{table}
The expressions of the amplitudes can be read off from Eq.~\eqref{eq:r1}-\eqref{eq:t1} and Eq.~\eqref{eq:r2}-\eqref{eq:t2}.  

\subsection{Model}
In this section, we introduce the Hamiltonian describing the environment, which will be the starting point for the scattering theory in the next section. 
\subsubsection{Ground- and inverted-state ferromagnet}
To describe the nonequilibrium environment, we use a generalized scattering theory in Bogoliubov space combining positive-energy and negative-energy magnonic excitations by using the language of antimagnonics~\cite{harms_2024}.  We follow the calculations in close analogy to Ref.~\cite{balbinot_2013}, where analog black holes in superfluids are studied. The full environment Hamiltonian can be decomposed as $H_\text{E}=\int\mathrm{d}x\left({\cal H }_\text{L}+{\cal H }_\text{R}+{\cal H }_\text{I}\right)$ which describes contributions of the left ground-state ferromagnet, the right inverted-state ferromagnet, and the coupling at the interface. 
We start with the bulk properties of the left and the right magnet. For this, we use an effective low-energy theory of the spin density $\vb{s}_\alpha$, where the index $\alpha\in\text{L,R}$  labels the left and right magnet, respectively. The Hamiltonian density takes the form
\begin{align}
{\cal H }_\alpha=  \frac{\tilde{A}}{2} \left(\partial_x \vb{s}_\alpha \right)^2  -\vb{h}_\alpha  \cdot \vb{s}_\alpha,
\end{align}
where  $\vb{s}_\text{L}$ and  $\vb{s}_\text{R}$ are only nonzero for $x<0$ and $x>0$, respectively. 
The first term describes the bulk exchange interaction with spin stiffness $\tilde{A}$ that favors a parallel alignment of the spins.  The second term describes the Zeeman energy, where we assume a stepwise magnetic field $\vb{h}_\text{L}=-\tilde{h}_\text{L} \vb{e}_z$ for $x<0$ and $\vb{h}_\text{R}=-\tilde{h}_\text{R}\vb{e}_z$ for $x>0$ with $\tilde{h}_\text{R}>\tilde{h}_\text{L}>0$.\footnote{Here, we define the field $\vb{h}_\text{L,R}$ as the actual magnetic field times the gyromagnetic ratio $\gamma_e=g_e \mu_\text{B}/\hbar<0$ with the g-factor $g_e\approx-2$ and the Bohr magneton $\mu_\text{B}$. Then, the energy is minimized when the spin density  $\vb{s}_\text{L,R}$  and the field $\vb{h}_\text{L,R}$  point in the same direction.} 
In the absence of ${\cal H }_\text{I}$,  the ferromagnet is aligned with $\vb{h}_\text{L}$ on the left side and, thus, in its ground state with $\ev{\vb{s}_\text{L}}=-s\vb{e}_z$, where $s$ is the spin density per length. On the right, on the other hand, the ferromagnet is antialigned with $\vb{h}_\text{R}$ which we refer to as the \textit{inverted state} with $\ev{\vb{s}_\text{R}}=+s\vb{e}_z$.
The equations of motion follow from Heisenberg equations and are given by
\begin{align}
\partial_t \vb{s}_\text{L} &= \vb{s}_\text{L}\times \left(\tilde{A} \partial_x^2 \vb{s}_\text{L} + \vb{h}_\text{L}  \right)   -\frac{\alpha}{s}  \vb{s}_\text{L} \times \partial_t \vb{s}_\text{L},  \quad \text{for} \quad x<0, \\
\partial_t \vb{s}_\text{R} &=  \vb{s}_\text{R}\times\left(\tilde{A} \partial_x^2 \vb{s}_\text{R} + \vb{h}_\text{R}  \right)   -\frac{\alpha}{s}  \vb{s}_\text{R} \times \partial_t \vb{s}_\text{R} +   \frac{ \tau_\text{S}}{s}\vb{s}_\text{R} \times\left(\vb{s}_\text{R} \times {\vb{e}_z} \right), \quad \text{for} \quad x>0, 
\end{align}
where we used the spin commutation relations $\left[ s_i(x),s_j(x^\prime)\right]=2\hbar \varepsilon_{ijk} s_k (x) \delta(x{-}x^\prime)$. In addition, we phenomenologically added dissipation via Gilbert damping with damping factor $\alpha$ and a spin torque described by $\tau_\text{S}$ to drive the system.

\begin{figure*}[h]
\includegraphics[width=16cm]{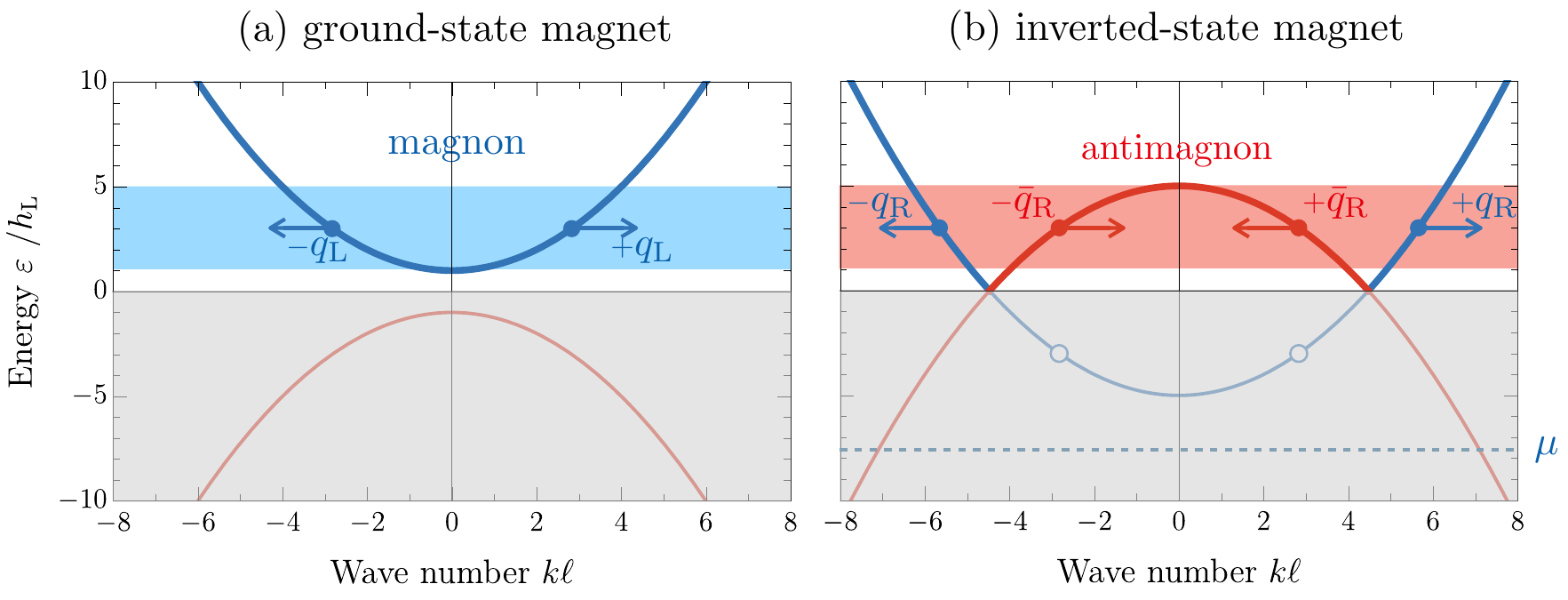}
	\caption{Magnon excitation energies $\varepsilon_\text{L,R}$ (blue lines) for (a) the left and (b) the right magnet. The right magnet is in an unstable inverted state, which gives rise to negative excitation energies, $\varepsilon_\text{R}<0$. Formally, we can add the antimagnonic excitation energies $\bar{\varepsilon}_\text{L,R}$ (red lines) with an inverted sign and disregard all negative energies (gray area). For energies $h_\text{L}<\varepsilon_\text{L}<h_\text{R}$ (blue area on the left)  and $h_\text{L}<\bar{\varepsilon}_\text{R}<h_\text{R}$ (red area on the right) magnons and antimagnons can be created pairwise at the interface. We choose $h_\text{R}=5 h_\text{L}$ and we define the length scale $\ell=\sqrt{\stiff/(h_\text{R}-h_\text{L})}$.}
	\label{fig:supp2}
\end{figure*}

For a linearized treatment of spin excitations, we use the usual Holstein-Primakoff bosonic fields in the large spin limit
\begin{align}
\psi_\text{L}&=\frac{1}{\sqrt{2\hbar s}}\left(s_\text{L}^x-is^y_\text{L}\right),\\
\psi_\text{R}&=\frac{1}{\sqrt{2\hbar s}}\left(s_\text{R}^x+is^y_\text{R}\right),
\end{align}
which describe magnonic excitations with spin $+\hbar$ and $-\hbar$ on top of the ground and the inverted state, respectively.  
They obey the bosonic commutation relations $\left[\psi_\text{L,R}(x),\psi^\dagger_\text{L,R}(x^\prime)\right]=\delta(x{-}x^\prime)$. 
The Hamiltonian becomes bilinear after normal ordering and we obtain
\begin{align}
{\cal H }_\alpha=\stiff \left(\partial_x \psi^\dagger_\alpha\right) \left(\partial_x \psi_\alpha\right)  \pm  {h}_\alpha  \psi^\dagger_\alpha \psi_\alpha,
\end{align}
where the positive (negative) sign holds for the left (right) magnet. Here, we defined a rescaled spin stiffness $A=\hbar s\tilde{A}/2$ and magnetic field  $h_\alpha=\hbar \tilde{h}_\alpha$. 
The  bulk field equations become 
\begin{align}
    i\hbar \partial_t {\psi}_\text{L}&=\left(1-i \alpha\right) \left(-\stiff  \partial_x^2+h_\text{L}\right){\psi}_\text{L}, \quad \text{for} \quad x<0, \\
     i\hbar \partial_t {\psi}_\text{R}&=\left(1-i \alpha\right) \left(-\stiff \partial_x^2-h_\text{R}\right){\psi}_\text{R}+  i \hbar \tau_\text{S} {\psi}_\text{R},  \quad \text{for} \quad x>0.
\end{align}
The associated bulk excitation energies of the magnons are given by $\varepsilon_\text{L,R}=\stiff k^2 \pm h_\text{L,R}$ (blue) for the left and right magnets which are shown in Fig.~\ref{fig:supp2}(a) and Fig.~\ref{fig:supp2}(b) as a function of the wave number $k$, respectively. The magnon energies are increased by $h_\text{L}$ in the left magnet, while they are lowered by $h_\text{R}$ in the right magnet. As a result, we obtain negative excitation energies, $\varepsilon_\text{R}<0$, which indicates the instability of the inverted magnet.
Due to Gilbert damping, the magnon excitation energies acquire an imaginary part, $-i \alpha \varepsilon$. For positive energies, $\varepsilon_\text{L,R}>0$, this leads to an exponential decay of the magnon mode. However, for negative energies, $\varepsilon_\text{L,R}<0$, the magnon modes grow exponentially and eventually flip the whole magnet to its ground state. Here, we want to avoid this growing instability by dynamically stabilizing the system with a spin torque described by $\tau_\text{S}$ that effectively suppresses all magnon excitations and leaves the system in the inverted vacuum state. Specifically,  if $-\hbar \tau_\text{s}>\alpha h_\text{R}$, the system will be dynamically stable. 

\subsubsection{Exchange coupling at interface}
At the interface, both magnets are coupled via exchange interaction 
\begin{align}
{\cal H }_\text{I}=\tilde{\gamma} \delta(x)\vb{s}_\text{L}\cdot \vb{s}_\text{R}\approx \gamma \delta(x) \left(\psi_\text{L}^\dagger \psi_\text{L}+\psi_\text{R}^\dagger \psi_\text{R}+\psi_\text{L} \psi_\text{R}+\psi_\text{L}^\dagger \psi_\text{R}^\dagger\right), \label{eq:intcoupling}
\end{align}
where we, again, rescale the coupling parameter through $\gamma=\hbar  s \tilde{\gamma}$.\footnote{Since $\psi_\text{L}(x)$ and $\psi_\text{L}(x)$ are only defined for $x\le 0$ and $x \ge 0$, respectively, the action of the delta function should be understood as  $\lim_{\varepsilon\rightarrow0^+}\int \mathrm{d} x\, \delta(x)\psi_\text{L}(x-\varepsilon) \psi_\text{R}(x+\varepsilon)=\psi_\text{L}(0)\psi_\text{R}(0)$, in accordance with the continuous limit of a discrete spin model. }
The sign of $\gamma$ can be either ferromagnetic or antiferromagnetic. Crucially, this coupling contains terms of the form $\sim   \delta(x) \psi_\text{L}^\dagger \psi_\text{R}^\dagger$ which become resonant when positive-energy magnons in the left magnet [blue filled dots in Fig.~\ref{fig:supp2}(a)] are matched with negative-energy magnons (or antimagnons) in the right magnet  [blue empty dots in Fig.~\ref{fig:supp2}(b)]. This can only happen for energies in the interval $h_\text{L}<\varepsilon_\text{L}<h_\text{R}$ (indicated by the blue area). In this way, the interfacial coupling serves as a controlled outlet to harvest the instability by creating entangled pairs of magnons.\footnote{We remark a formal similarity to a parametric drive often used in quantum optics to create squeezed states. Here, however, ${\cal H}_\text{C}$ is not a drive but an energy-conserving coupling Hamiltonian that merely connects the stable and inverted region with each other and, thereby, enables the pair creation. Here, the instability is disspatively stabilized. } 
The exchange coupling modifies the equations of motion to 
\begin{align}
    i\hbar \partial_t 
   {\psi}_\text{L} &= \left[-\stiff \partial_x^2+h_\text{L} \right]    {\psi}_\text{L} +  \gamma \delta(x)\left( {\psi}_\text{L}+ {\psi}_\text{R}^\dagger\right), \label{eq:fieldboundary1} \\
       i\hbar \partial_t 
   {\psi}_\text{R} &= \left[-\stiff \partial_x^2-h_\text{R} \right]    {\psi}_\text{R} +  \gamma \delta(x)\left( {\psi}_\text{R}+ {\psi}_\text{L}^\dagger\right),
\label{eq:fieldboundary2}\end{align}
where we neglect dissipation and driving in the following. Note that creation and annihilation operators are coupled at the interface.
 Therefore, in the next section, we will extend the formalism to the bosonic Bogoliubov-de-Gennes space.

\subsection{Scattering theory}
Our goal is to find the magnon field operator $\psi_\alpha(x,t)$ expanded in the scattering solutions of the problem. It can be written using the Bogoliubov ansatz~\cite{balbinot_2013,harms_2024}
\begin{align}
\psi_\alpha(x,t)= \int \limits_0^\infty\mathrm{d}\varepsilon \sum_i\Bigg[e^{-i \varepsilon t/\hbar}u_{i}(\varepsilon, x) a_i({\varepsilon})+e^{i\varepsilon t/\hbar}v^*_{i}(\varepsilon, x) a^\dagger _{i}(\varepsilon)\Bigg], \label{eq:bogansatz}
\end{align}
where $u_{i}$  and $v_{i}$ correspond to the normal and anomalous amplitudes of the scattering solutions, respectively. Here, the operators $a_i$ describe the inscattering eigenstates (defined below) at a given energy $\varepsilon$ and hence diagonalize the scattering problem. They fulfill the commutation relations 
\begin{align}
\left[a_i(\varepsilon),a^\dagger_j(\varepsilon^\prime)\right]= \Sigma_{ij} \delta(\varepsilon{-}\varepsilon^\prime), \label{eq:a_commutator}
\end{align}
where the diagonal signature matrix is given by $\Sigma_{ij}={\nu_i} \delta_{ij}$. 
Due to the restriction to positive energies, $\varepsilon>0$, we obtain two kinds of inscattering solutions with $\nu_i=\pm1$. The sign is determined by the character of the incoming state. For an incoming positive-energy magnon, we get ${\nu}_i=+1$, and the operator $a_i(\varepsilon)$ functions as an annihilation operator. For an incoming negative-energy magnon (or antimagnon), we get ${\nu}_i=-1$, and the operator $a_i(\varepsilon)=b_i^\dagger(-\varepsilon)$ functions as a creation operator.  Formally, the sign is determined by the normalization of the scattering modes~\cite{balbinot_2013}:
\begin{align}
\int\mathrm{d} x\left[u^*_i(\varepsilon, x) u_j(\varepsilon^\prime, x)-v_i(\varepsilon, x) v_j^*(\varepsilon^\prime, x) \right]=\nu_i \delta_{ij} \delta(\varepsilon-\varepsilon^\prime).
\end{align}
The negative-norm states correspond to the  negative-energy states induced by the instability.

\subsubsection{Classical scattering problem}
To reduce the field operator equations from Eq.~\eqref{eq:fieldboundary1}-\eqref{eq:fieldboundary2}  to a classical scattering problem, we insert Eq.~\eqref{eq:bogansatz} and apply the commutator from Eq.~\eqref{eq:a_commutator} to obtain
\begin{align}
  \varepsilon\,\underline{\varphi}_{\text{L},i} &={\cal H}_\text{L}^\text{BdG}  \underline{\varphi}_{\text{L},i } 
, \quad \text{for} \quad x<0, \label{eq:bulkL} \\
\varepsilon  \, \underline{\varphi}_{\text{R},i} &={\cal H}_\text{R}^\text{BdG}  \underline{\varphi}_{\text{R},i}, \quad \text{for} \quad x>0, \label{eq:bulkR}
 \end{align}
where we introduce the classical field $\underline{\varphi}_{\alpha,i}=\left(u_{\alpha,i},v_{\alpha,i}\right)$ with the indices $\alpha=\text{L}$ and $\alpha=\text{R}$ to indicate $x<0$ and $x>0$, respectively. Here, we defined the effective bosonic  Bogoliubov-de-Gennes Hamiltonian  ${\cal H}_{\text{L,R}}^\text{BdG}=\left(-\stiff \partial_x^2 \pm h_\text{L,R}\right)\tau_z$,  where $\tau_x$, $\tau_y$, and $\tau_z$ are the Pauli matrices.\footnote{We remark that even though the Hamiltonians $ {\cal H}^\text{BdG}_\alpha$ are hermitian, hermiticity will not be guaranteed for more generic systems when for example a magnetic anisotropy is included.
However, the Hamiltonian will always obey pseudo hermiticity,  $\tau_z {\cal H}_\alpha^\text{BdG}   \tau_z = \left({\cal H}_\alpha^\text{BdG}\right)^\dagger$, as the evolution generated by  ${\cal H}_\alpha^\text{BdG}$ conserves the commutation relations of $\psi_\alpha$ and $\psi_\alpha^\dagger$. } 
The corresponding bulk eigenvalues are given by
\begin{align}
\varepsilon_\text{L,R}=\stiff k^2 \pm h_\text{L,R}, \quad \bar{\varepsilon}_\text{L,R}=-\stiff k^2 \mp h_\text{L,R},
\end{align}
which are both shown in Fig.~\ref{fig:supp2} in blue and red, respectively. Since we effectively introduced a redundancy by doubling the degrees of freedom, positive- and negative-energy solutions are related to each other through~\cite{harms_2024}
\begin{align}
\tau_x {\cal H}_\alpha^\text{BdG}   \tau_x = -\left({\cal H}_\alpha^\text{BdG}\right)^*.
\end{align} 
This justifies that we can disregard all negative energies (gray area in Fig.~\ref{fig:supp2}) and only consider $\varepsilon>0$. For the instability on the right side, $\varepsilon_\text{R}<0$, this means that negative-energy magnons (or antimagnons) will be described by the inverted (or antimagnonic) branch $\bar{\varepsilon}_\text{R}$. 

Furthermore, the coupling at the interface leads to the boundary conditions 
\begin{align}
   \left(\partial_x + \frac{\gamma}{A}\right)\underline{\varphi}_{\text{L},i}
    &=- \frac{\gamma}{A}\,\tau_x    \underline{\varphi}_{\text{R},i}, \quad \text{for} \quad x=0,  \label{eq:boundary1} \\
\left(\partial_x- \frac{\gamma}{A}\right)
   \underline{\varphi}_{\text{R},i}
    &=+ \frac{\gamma}{A}\,\tau_x   \underline{\varphi}_{\text{L},i}, \quad \text{for} \quad x=0, \label{eq:boundary2} \end{align}
which can be obtained by integrating Eq.~\eqref{eq:fieldboundary1}-\eqref{eq:fieldboundary2} over a small interval around the interface $x=0$. 
Due to $\tau_x$, we only couple normal modes on the left ($u_{\text{L},i}$) with anomalous modes on the right  ($v_{\text{R},i}$)  and vice versa.  

\subsubsection{Ansatz for scattering solution}
To find the scattering solutions, we use the ansatz
\begin{align}
   \underline{\varphi}_{\alpha,i}&=   a_{\alpha,i}  \underline{m}_{\alpha,+} + b_{\alpha,i}\underline{m}_{\alpha,-}+   \bar{a}_{\alpha,i} \bar{\underline{m}}_{\alpha,+} + \bar{b}_{\alpha,,i}\bar{\underline{m}}_{\alpha,-},  \label{eq:ansatzscatteringfunction}
\end{align}
where $a_{\alpha,i}, \bar{a}_{\alpha,i}, b_{\alpha,i}$ and $\bar{b}_{\alpha,i}$  are the energy-dependent scattering amplitudes and $i$ labels the different scattering solutions. Here, $\underline{m}_{\alpha, \pm}$ and $\bar{\underline{m}}_{\alpha, \pm}$ are the four eigenmodes of the bulk equations Eq.~\eqref{eq:bulkL}-\eqref{eq:bulkR}, where $\alpha=\text{L}$ denotes $x<0$ and $\alpha=\text{R}$ denotes $x>0$.  They are given by
\begin{align}
\underline{m}_{\alpha, \pm} = \frac{1}{\sqrt{2\pi\hbar v_\alpha}} e^{\pm i q_\alpha x}  \begin{pmatrix}
   1\\
  0
\end{pmatrix}, \quad  \bar{\underline{m}}_{\alpha, \pm} =\frac{1}{\sqrt{2\pi\hbar \bar{v}_\alpha}} e^{\mp i\bar{q}_\alpha x} \begin{pmatrix}
   0\\
 1
\end{pmatrix},\label{eq:planewaves}
\end{align}
where the first term describes pure right-moving ($\underline{m}_{\alpha,+}$) and left-moving ($\underline{m}_{\alpha,-}$) magnons (from the regular dispersion) and the second term describes pure right-moving ($\bar{\underline{m}}_{\alpha,+}$)  and left-moving ($\bar{\underline{m}}_{\alpha,-}$) antimagnons (from the inverted dispersion). Since the spin $z$ angular momentum is conserved, the eigenmodes do not mix normal and anomalous amplitudes. The wave numbers and absolute group velocities are denoted by $q_\alpha$ and ${v}_\alpha$ for magnons and $\bar{q}_\alpha$ and $\bar{v}_\alpha$ for antimagnons, respectively. Note that for antimagnons, we introduced a relative sign change in the exponent since momentum and group velocity are of opposite sign.  The wave numbers  in the left magnet can be derived from Eq.~\eqref{eq:bulkL} and are given by
\begin{align}
    q_\text{L} =  \frac{1}{\sqrt{A}}\begin{cases}
    {\sqrt{\varepsilon-h_\text{L}}}, & \text{for } \varepsilon \geq h_\text{L} \\
    i {\sqrt{h_\text{L}-\varepsilon}}, & \text{for } \varepsilon \leq h_\text{L} 
  \end{cases} \quad \text{and} \quad 
 \bar{q}_\text{L} = -i \frac{\sqrt{\varepsilon+h_\text{L}}}{\sqrt{A}}.
\end{align}
The wave numbers in the right magnet are derived from Eq.~\eqref{eq:bulkR} and are given by
\begin{align}
    q_\text{R}=\frac{\sqrt{\varepsilon+h_\text{R}}}{\sqrt{A}}  \quad \text{and} \quad 
    \bar{q}_\text{R} =  \frac{1}{\sqrt{A}}\begin{cases}
     -i {\sqrt{\varepsilon-h_\text{R}}},   & \text{for } \varepsilon \geq h_\text{R} \\
{\sqrt{h_\text{R}-\varepsilon}},  & \text{for } \varepsilon \leq h_\text{R} 
  \end{cases}.
\end{align}
We choose the signs of the imaginary wave numbers such that the derived results below do not contain exponentially growing solutions.  
The absolute values of the group velocities are given by
\begin{align}
    v_\alpha=\frac{2\stiff \abs{q_\text{L}}}{\hbar}, \quad \bar{v}_\alpha =\frac{2\stiff \abs{\bar{q}_\text{L}}}{\hbar},
\end{align}
which are only meaningful for propagating modes. However, to keep the notation uniform, we also define them for the evanescent modes. They are included in Eq.~\eqref{eq:planewaves} to ensure a proper normalization in the energy eigenbasis.\footnote{By including the group velocity, the propagating eigenmodes are normalized on delta functions in energy,  $\int\mathrm{d}x  \frac{1}{\sqrt{2\pi\hbar v_\alpha(\varepsilon)}}e^{+i q_\alpha(\varepsilon)  x} \frac{1}{\sqrt{2\pi\hbar v_\alpha(\varepsilon^\prime)}}e^{-i q_\alpha(\varepsilon^\prime)  x}=\delta(\varepsilon-\varepsilon^\prime)$.}

\subsubsection{Matching of scattering amplitudes}
To obtain the energy-dependent scattering amplitudes $a_{\alpha,i}, \bar{a}_{\alpha,i}, b_{\alpha,i}$ and $\bar{b}_{\alpha,i}$ of the scattering solution Eq.~\eqref{eq:ansatzscatteringfunction}, we use the boundary condition to link the amplitudes on the left and right side of the interface via 
\begin{align}
 \begin{pmatrix}
   a_{\text{L},i}\\
  b_{\text{L},i}\\
  \bar{a}_{\text{L},i}\\
  \bar{b}_{\text{L},i}
\end{pmatrix}= {\cal M} \begin{pmatrix}
   a_{\text{R},i}\\
  b_{\text{R},i}\\
  \bar{a}_{\text{R},i}\\
  \bar{b}_{\text{R},i}
\end{pmatrix}. \label{eq:matching}
\end{align}
The transfer matrix is defined by ${\cal M}=\left({\cal W}_\text{L} {\cal N}_\text{L}\right)^{-1}{\cal W}_\text{R} {\cal N}_\text{R}$, where 
we introduced the diagonal normalization matrices ${\cal N}_\alpha= \sqrt{2\pi \,\text{diag}\left(v_\alpha,v_\alpha,\bar{v}_\alpha,\bar{v}_\alpha\right)}^{-1}$ as well as the matrices
\begin{align}
{\cal W}_\text{L}=
\begin{pmatrix}
   i q_\text{L} +\frac{\gamma}{A} &  -i q_\text{L} +\frac{\gamma}{A} &0&0 \\
     0&0&-i \bar{q}_\text{L} +\frac{\gamma}{A} &  i \bar{q}_\text{L} +\frac{\gamma}{A}  \\
       \frac{\gamma}{A}&\frac{\gamma}{A}& 0 &0 \\
      0&0&\frac{\gamma}{A} &\frac{\gamma}{A} 
\end{pmatrix}, \quad 
{\cal W}_\text{R}=
\begin{pmatrix}
   0&0&-\frac{\gamma}{A} &-\frac{\gamma}{A} \\
     -\frac{\gamma}{A} &-\frac{\gamma}{A} &0&0 \\
          0&0&-i \bar{q}_\text{R} -\frac{\gamma}{A} & +i \bar{q}_\text{R} -\frac{\gamma}{A}  \\
   i q _\text{R}-\frac{\gamma}{A} &  -i q_\text{R} -\frac{\gamma}{A} &0&0 
\end{pmatrix},
\end{align}
which directly follow from Eq.~\eqref{eq:boundary1}-\eqref{eq:boundary2} and Eq.~\eqref{eq:ansatzscatteringfunction}.

\subsubsection{Inscattering states}
The matching conditions given by Eq.~\eqref{eq:matching} have to be complemented with scattering boundary conditions. 
Here, we construct the inscattering states $\underline{\varphi}_{\alpha,i}^\text{in}$ of the problem, where a plane-wave mode $\underline{m}_{\alpha,\pm}$ or  $\bar{\underline{m}}_{\alpha,\pm}$ with amplitude $1$ is impinging on the interface and gets scattered into the outgoing plane-wave modes. 

The number of possible inscattering solutions is energy dependent. For  $h_\text{L}<\varepsilon<h_\text{R}$, we have the three inscattering solutions, where  $\underline{m}_{\text{L},+}$ (magnon from the left), $\underline{m}_{\text{R},-}$ (magnon from the right), and $\bar{\underline{m}}_{\text{R},-}$ (antimagnon from the right) are moving towards the interface, see Fig.~\ref{fig:supp2}. For  $0<\varepsilon<h_\text{L}$, the mode $\underline{m}_{\text{L},+}$ becomes evanescent, reducing the number of scattering solutions to 2. Similarly, for $\varepsilon>h_\text{R}$, the mode  $\bar{\underline{m}}_{\text{R},-}$ becomes evanescent, reducing the number of scattering solutions to 2. 
In the following, we will consider all scattering solutions in the interval $h_\text{L}<\varepsilon<h_\text{R}$. However, the results are written such that they are correct for all energies.

\paragraph{\textbf{Incoming magnon from the left:}}\label{sec:sol1}
\begin{figure}[h]
\includegraphics[width=13cm]{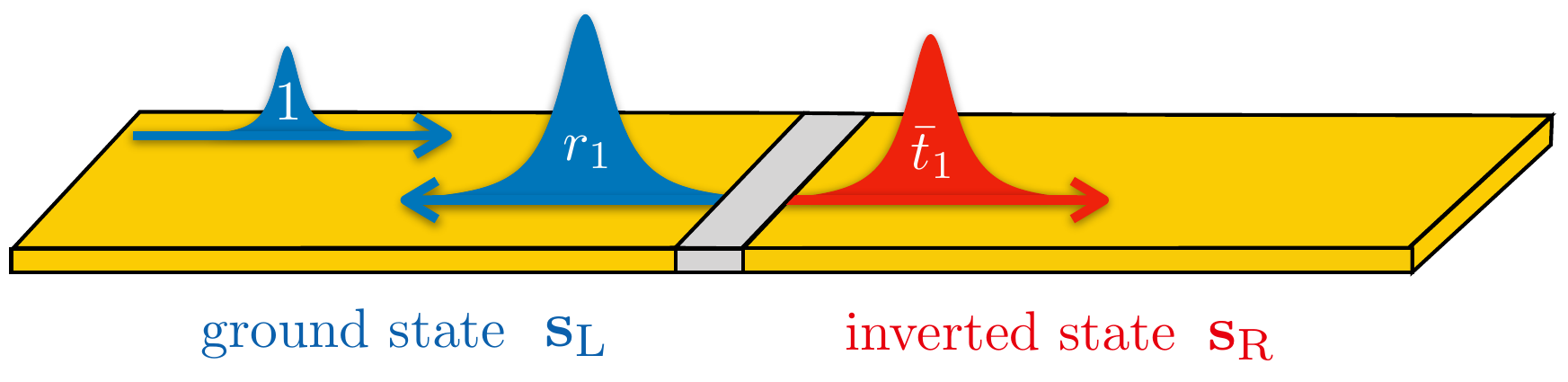}
	\caption{Schematics of the inscattering solution. An incoming magnon with amplitude $1$ is reflected with an enhanced amplitude $\abs{r_1}>1$ and transmitted as an antimagnon with $\bar{t}_1$. The relative size of the wave packets schematically indicates that $\abs{r_1}^2-\abs{\bar{t}_1}^2=1$.}
	\label{fig:supp3}
\end{figure}
To find the first inscattering solution $\underline{\varphi}^\text{in}_{\alpha,1}$, see Fig.~\ref{fig:supp3}, we assume for the scattering boundary condition an incoming magnon mode from the left. We use the ansatz
\begin{align}
\begin{pmatrix}
   a_{\text{L},1}\\
  b_{\text{L},1}\\
  \bar{a}_{\text{L},1}\\
  \bar{b}_{\text{L},1}
\end{pmatrix}= \begin{pmatrix}
  i_1\\
 r_1\\
  0\\
 \bar{r}_1
\end{pmatrix}, \quad 
 \begin{pmatrix}
   a_{\text{R},1}\\
  b_{\text{R},1}\\
  \bar{a}_{\text{R},1}\\
  \bar{b}_{\text{R},1}
\end{pmatrix}= \begin{pmatrix}
  t_1\\
0\\
 \bar{t}_1\\
  0
\end{pmatrix}. 
\end{align}
There is an incoming right-moving magnon from the left with amplitude 
\begin{align}
i_1=\Theta(\varepsilon-h_\text{L})
\end{align}
that can be reflected back with amplitude $r_1$ as a propagating magnon, or it can be reflected  with amplitude $\bar{r}_1$ as an evanescent antimagnon. On the right side, the magnon can be transmitted with amplitude $t_1$ as a propagating magnon, or it can be transmitted with amplitude  $\bar{t}_1$ as a propagating (evanescent) antimagnon for $\varepsilon<h_R$ ($\varepsilon>h_R$). The remaining amplitudes are zero because no other propagating modes are incoming, and we disregard exponentially growing solutions. 
With Eq.~\eqref{eq:matching}, we find for the magnonic reflection amplitude
\begin{align}
r_1= \Theta(\varepsilon-h_\text{L})\frac{q_\text{L} \bar{q}_\text{R} - i \frac{\gamma}{A} \left(q_\text{L}+ \bar{q}_\text{R} \right)}{q_\text{L} \bar{q}_\text{R} - i \frac{\gamma}{A} \left(q_\text{L}- \bar{q}_\text{R}\right)},  \label{eq:r1}
\end{align}
and for the antimagnonic transmission amplitude
\begin{align}
 \bar{t}_1= \Theta(\varepsilon-h_\text{L})\frac{-2 \frac{\gamma}{A} \sqrt{\abs{{\bar{q}_\text{R}}{{q}_\text{L}}}} }{i q_\text{L} \bar{q}_\text{R} +  \frac{\gamma}{A} \left(q_\text{L}- \bar{q}_\text{R}\right)}. \label{eq:t1}
\end{align}
We also obtain $\bar{r}_1=t_1=0$, because the spin $z$ angular momentum is conserved. Hence, the magnon can neither be reflected as an antimagnon nor can it be transmitted as a magnon. Note that 
\begin{align}
\abs{{r}_1}^2-\abs{ \bar{t}_1}^2=1 \quad  \text{for} \quad h_\text{L}< \varepsilon<h_\text{R}.
\end{align}
The reflection of magnons is effectively enhanced, $\abs{{r}_1}^2\geq1$, which is already a hint at the underlying pair creation of magnons and antimagnons at the interface.  

\paragraph{\textbf{Incoming antimagnon from the right:}}\label{sec:sol2}
\begin{figure}[h]
\includegraphics[width=13cm]{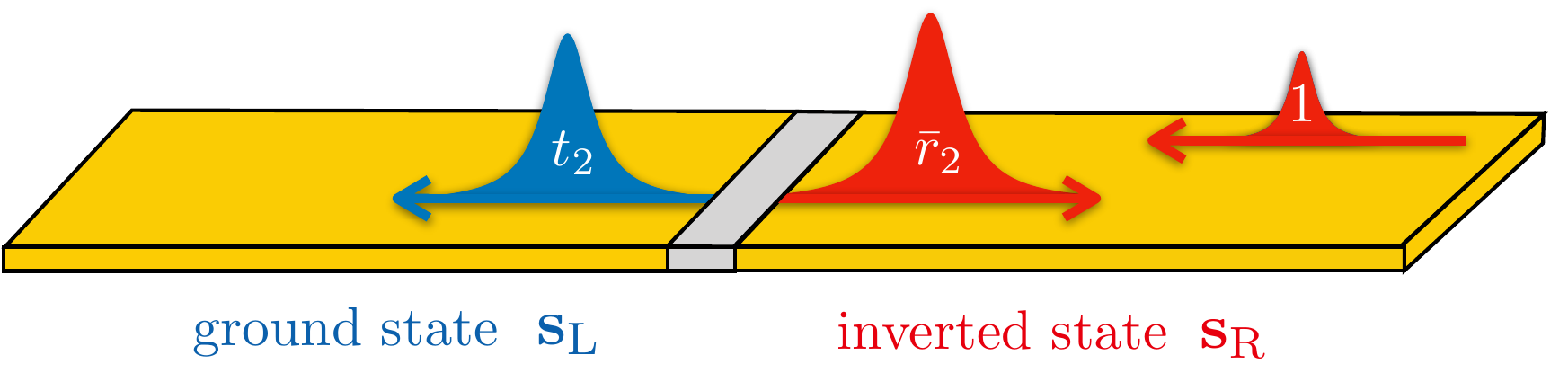}
	\caption{Schematics of the inscattering solution. An incoming antimagnon with amplitude $1$ is reflected with an enhanced amplitude $\abs{\bar{r}_2}>1$ and transmitted as a magnon with ${t}_2$.  The relative size of the wave packets schematically indicates that $\abs{\bar{r}_2}^2-\abs{{t}_2}^2=1$.}
	\label{fig:supp4}
\end{figure}
To find the second inscattering solution $\underline{\varphi}^\text{in}_{\alpha,2}$, see Fig.~\ref{fig:supp4}, we assume for the scattering boundary condition an incoming antimagnon from the right. We use the ansatz 
\begin{align}
\begin{pmatrix}
   a_{\text{L},2}\\
  b_{\text{L},2}\\
  \bar{a}_{\text{L},2}\\
  \bar{b}_{\text{L},2}
\end{pmatrix}= \begin{pmatrix}
  0\\
 t_2\\
  0\\
 \bar{t}_2
\end{pmatrix}, \quad 
 \begin{pmatrix}
   a_{\text{R},2}\\
  b_{\text{R},2}\\
  \bar{a}_{\text{R},2}\\
  \bar{b}_{\text{R},2}
\end{pmatrix}= \begin{pmatrix}
  r_2\\
  0\\
 \bar{r}_2\\
  \bar{i}_2
\end{pmatrix}. 
\end{align}
There is an incoming left-moving antimagnon from the right with amplitude 
\begin{align}
\bar{i}_2=\Theta(h_\text{R}-\varepsilon)
\end{align}
that can be reflected back with amplitude $\bar{r}_2$ as a propagating antimagnon,  or it can be reflected with amplitude ${r}_2$ as a propagating magnon. On the left side, the antimagnon can be transmitted  with amplitude $\bar{t}_2$ as an evanescent antimagnon, or it can be transmitted with amplitude $t_2$ as a propagating (evanescent) magnon for $\varepsilon>h_\text{L}$ ($\varepsilon<h_\text{L}$). The remaining amplitudes are zero because no other propagating modes are incoming, and we disregard exponentially growing solutions. 
With Eq.~\eqref{eq:matching}, we find for the antimagnonic reflection amplitude
\begin{align}
\bar{r}_2= \Theta(h_\text{R}-\varepsilon)\frac{q_\text{L} \bar{q}_\text{R} + i \frac{\gamma}{A} \left(q_\text{L}+ \bar{q}_\text{R} \right)}{q_\text{L} \bar{q}_\text{R} - i \frac{\gamma}{A} \left(q_\text{L}- \bar{q}_\text{R}\right)},  \label{eq:r2}
\end{align}
and for the magnonic transmission amplitude
\begin{align}
 {t}_2=\Theta(h_\text{R}-\varepsilon)\frac{2 \frac{\gamma}{A} \sqrt{\abs{{{q}_\text{L}}{\bar{q}_\text{R}}}} }{i q_\text{L} \bar{q}_\text{R} +  \frac{\gamma}{A} \left(q_\text{L}- \bar{q}_\text{R}\right)}. \label{eq:t2}
\end{align}
We also obtain ${r}_2=\bar{t}_2=0$, because the spin $z$ angular momentum is conserved. Hence, the antimagnon can neither be reflected as a magnon nor can it be transmitted as an antimagnon. Note that 
\begin{align}
\abs{\bar{r}_2}^2-\abs{ {t}_2}^2=1 \quad  \text{for} \quad h_\text{L}< \varepsilon<h_\text{R}.
\end{align}
The reflection of antimagnons is effectively enhanced, $\abs{{r}_1}^2\geq1$, which is already a hint at the underlying pair creation of magnons and antimagnons at the interface.  
 
\paragraph{\textbf{Incoming magnon from the right:}}\label{sec:sol3}
\begin{figure}[h]
\includegraphics[width=13cm]{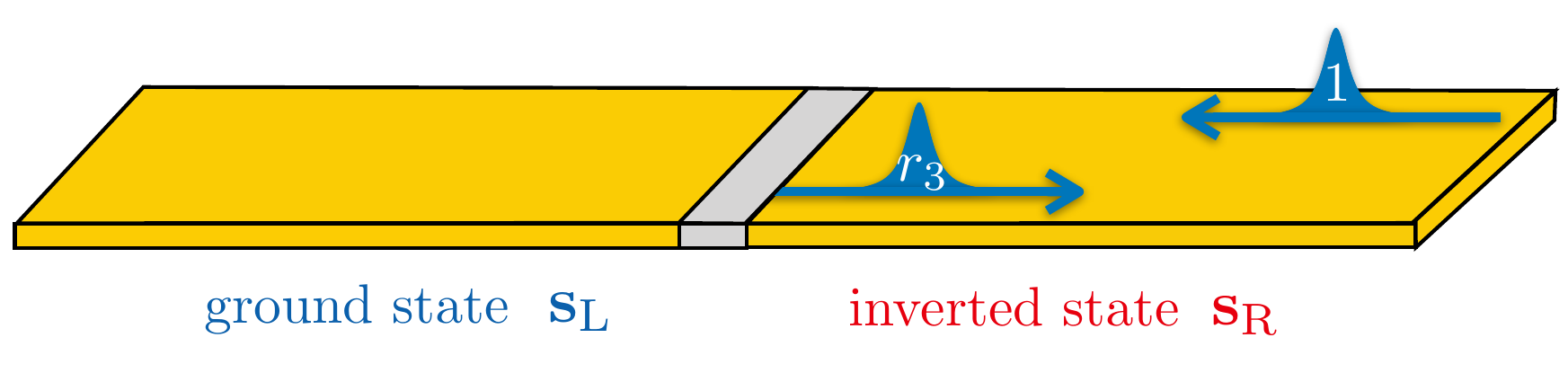}
	\caption{Schematics of the inscattering solution. An incoming magnon with amplitude $1$ is fully reflected with amplitude $\abs{{r}_3}=1$ and there is no transmission (only evanescent modes).  }
	\label{fig:supp5}
\end{figure}
To find the third inscattering solution $\underline{\varphi}^\text{in}_{\alpha,3}$, see Fig.~\ref{fig:supp5}, we assume for the scattering boundary condition an incoming magnon mode from the right. We use the ansatz 
\begin{align}
\begin{pmatrix}
   a_{\text{L},3}\\
  b_{\text{L},3}\\
  \bar{a}_{\text{L},3}\\
  \bar{b}_{\text{L},3}
\end{pmatrix}= \begin{pmatrix}
  0\\
 t_3\\
  0\\
 \bar{t}_3
\end{pmatrix}, \quad 
 \begin{pmatrix}
   a_{\text{R},3}\\
  b_{\text{R},3}\\
  \bar{a}_{\text{R},3}\\
  \bar{b}_{\text{R},3}
\end{pmatrix}= \begin{pmatrix}
  r_3\\
  i_3\\
 \bar{r}_3\\
  0
\end{pmatrix}. 
\end{align}
There is an incoming left-moving magnon from the right with amplitude 
\begin{align}
i_3=1
\end{align}
than can be reflected back with amplitude  ${r}_3$ as a propagating magnon, or it can be reflected with amplitude $\bar{r}_3$ as a propagating (evanescent) antimagnon for $\varepsilon<h_\text{R}$ ($\varepsilon>h_\text{R}$). On the left side, the magnon can be transmitted with amplitude $t_3$ as a propagating (evanescent) magnon for $\varepsilon>h_\text{L}$ ($\varepsilon<h_\text{L}$), or it can be transmitted with amplitude $\bar{t}_3$ as an evanescent antimagnon. The remaining amplitudes are zero because no other propagating modes are incoming, and we disregard exponentially growing solutions. 
With Eq.~\eqref{eq:matching}, we find for the magnonic reflection amplitude
\begin{align}
{r}_3= \frac{\bar{q}_\text{L} {q}_\text{R} - i \frac{\gamma}{A} \left(\bar{q}_\text{L}+ {q}_\text{R} \right)}{\bar{q}_\text{L} {q}_\text{R} +i \frac{\gamma}{A} \left(\bar{q}_\text{L}- {q}_\text{R}\right)}, 
\end{align}
and for the evanescent antimagnonic amplitude on the left side
\begin{align}
 \bar{t}_3= \frac{-2 \frac{\gamma}{A} \sqrt{\abs{\bar{q}_\text{L} {q}_\text{R}}} }{i \bar{q}_\text{L} {q}_\text{R} -  \frac{\gamma}{A} \left(\bar{q}_\text{L}- {q}_\text{R}\right)}.
\end{align}
We also obtain $\bar{r}_3={t}_3=0$, because the spin $z$ angular momentum is conserved. Hence, the magnon can neither be reflected as an antimagnon nor can it be transmitted as a magnon. Note that $\abs{r_3}^2=1$, i.e., the magnon is fully reflected. 

\subsection{Magnon field operators}
Inserting the amplitudes of the three inscattering states from Sec.~\ref{sec:sol1}- \ref{sec:sol3} into the ansatz of Eq.~\eqref{eq:ansatzscatteringfunction}, we find with Eq.~\eqref{eq:bogansatz} the magnon field operators
\begin{align}
{\psi}_\text{L}(x,t)&=\int \limits_{0}^{\infty}\frac{\mathrm{d}\varepsilon}{\sqrt{2\pi \hbar  v_\text{L}} } \Bigg[\left(i_1 e^{i q_\text{L} x}+ r_1 e^{-i q_\text{L}  x}\right)e^{-i\varepsilon t/\hbar}a_{1} + t_2 e^{-i q_\text{L} x}e^{-i\varepsilon t/\hbar}a_{2} +\sqrt{\frac{v_\text{L}}{\bar{v}_\text{L}}}\left(\bar{t}_3 e^{i\bar{q}_\text{L} x}\right)^* e^{i\varepsilon t/\hbar}a^\dagger_{3}\Bigg],\\
{\psi}^\dagger_\text{R}(x,t)&=\int \limits_{0}^{\infty}\frac{\mathrm{d}\varepsilon}{\sqrt{2\pi \hbar  \bar{v}_\text{R}} } \Bigg[  \left(\bar{t}_1 e^{-i \bar{q}_\text{R}  x}\right)e^{-i\varepsilon t/\hbar}a_{1} +\left(\bar{i}_2 e^{i \bar{q}_\text{R} x} + \bar{r}_2 e^{-i \bar{q}_\text{R} x}\right)e^{-i\varepsilon t/\hbar}a_{2} +\sqrt{\frac{\bar{v}_\text{R}}{{v}_\text{R}}}\left(i_3 e^{-i{q}_\text{R} x}+r_3 e^{i{q}_\text{R} x}\right)^* e^{i\varepsilon t/\hbar}a^\dagger_{3}\Bigg],
\end{align}
where the first (second) line holds for $x<0$ ($x>0$). In the following, we will be interested in the vacuum state $\ket{\Omega}$, where no magnon or antimagnon is incoming from the boundary. It can be defined through
\begin{align}
a_1 \ket{\Omega}=a_2^\dagger \ket{\Omega}=a_3 \ket{\Omega}=0,
\end{align}
for each energy $\varepsilon$. Note that for the antimagnonic mode, we have  $a_2(\varepsilon)=b^\dagger_2(-\varepsilon)$. Thus, it effectively corresponds to the creation operator of a negative-energy magnon. 

\subsection{Spin current}
Having solved for the magnon field operators ${\psi}_\text{L}(x,t)$ and ${\psi}_\text{R}(x,t)$, we can evaluate  the spin current via
\begin{align}
I^s_\alpha=- :\tilde{A} \,\vb{e}_z \cdot \left( \vb{s}_\alpha \times \partial_x \vb{s}_\alpha\right):,
\end{align}
where we use normal ordering in $\psi_\alpha$ and $\psi_\alpha^\dagger$. It results in
\begin{align}
    {{I}}^{s}_\text{L}&= +2\stiff \left[{\psi}_\text{L}^\dagger \left(-i \partial_x {\psi}_\text{L}\right)-\left(-i\partial_x {\psi}_\text{L}^\dagger\right) {\psi}_\text{L}\right],\quad \text{for} \quad x<0, \\
      {{I}}^{s}_\text{R}&=- 2\stiff \left[{\psi}_\text{R}^\dagger \left(-i \partial_x {\psi}_\text{R}\right)-\left(-i\partial_x {\psi}_\text{R}^\dagger\right)  {\psi}_\text{R}\right]  , \quad \text{for} \quad x>0.
\end{align}
In the vacuum $\ket{\Omega}$,  we get a finite spin current
\begin{align}
\ev{ {{I}}^{s}_\text{L}(x)}{\Omega}=-\frac{2\hbar}{ h} \int \limits_{h_\text{L}}^{h_\text{R}}{\mathrm{d}\en} \,\vert t_2(\en) \vert^2, \\
\ev{ {{I}}^{s}_\text{R}(x)}{\Omega}=-\frac{2 \hbar}{h} \int \limits_{h_\text{L}}^{h_\text{R}}{\mathrm{d}\en} \,\vert \bar{t}_1(\en) \vert^2,
\end{align}
where we write the proportionality constant in analogy to the electron current in the Landauer-Büttiker formalism~\cite{lesovik_2011}, where we get $\frac{2e}{ h}$ with $h=2\pi\hbar$.   Thus, the system sustains a spin current that is moving to the left. This current is purely quantum as it originates in the vacuum fluctuations at the interface. It manifests itself as a constant creation of magnon-antimagnon pairs, where $+\hbar$ magnons move into the left magnet and  $-\hbar$ antimagnons move into the right magnet. The spin current is the same for all $x$ since $\abs{\bar{t}_1}=\abs{t_2}=t$. We can also define the spin current per energy via $j^s(\varepsilon)= \,t^2/\pi$.

\subsection{Magnon-antimagnon pair creation}
To describe the pair creation, we first introduce the nonunitary scattering matrix $S$ in Sec.~\ref{sec:scatteringmatrix} to relate inscattering and outscattering states. Then, we use the scattering matrix $S$ to identify two-mode squeezing in Sec.~\ref{sec:squeezing} and we determine the probability of finding $n$ magnon-antimagnon pairs in the invacuum state $\ket{\Omega}$.
\subsubsection{Scattering matrix}\label{sec:scatteringmatrix}
To define the scattering matrix $S$, we write the three scattering solutions from Sec.~\ref{sec:sol1}- \ref{sec:sol3} via
\begin{align}
\begin{pmatrix}
   \underline{\varphi}^\text{in}_{1}\\
   \underline{\varphi}^\text{in}_{2}\\
  \underline{\varphi}^\text{in}_{3}
\end{pmatrix}= \begin{pmatrix}
  \underline{m}_{\text{L},+}\\
  \bar{\underline{m}}_{\text{R},-}\\
 {\underline{m}}_{\text{R},-}
\end{pmatrix} 
+   \underbrace{\begin{pmatrix}
  r_1 & \bar{t}_1 & 0\\
 t_2 & \bar{r}_2 & 0\\
 0&0&r_3
\end{pmatrix}}_{=S^T} \begin{pmatrix}
  \underline{m}_{\text{L},-}\\
  \bar{\underline{m}}_{\text{R},+}\\
 {\underline{m}}_{\text{R},+}
\end{pmatrix},\label{eq:scattdef1}
\end{align}
where we used the notation $\underline{\varphi}^\text{in}_i=\Theta(-x)\underline{\varphi}^\text{in}_{\text{L},i} +\Theta(x)\underline{\varphi}^\text{in}_{\text{R},i}$, and the scattering matrix $S^T$ summarizes how the incoming plane-wave modes are scattered into outgoing plane-wave modes. 
Furthermore, we assumed $h_\text{L}<\varepsilon<\varepsilon_\text{R}$ and neglected the evanescent modes by employing the asymptotic limit $x\rightarrow \pm \infty$. 

Alternatively, the $S$ matrix can be defined as the linear relation between inscattering states $\underline{\varphi}^\text{in}_i$ and outscattering states $\underline{\varphi}^\text{out}_i$,
\begin{align}
\underline{\varphi}^\text{in}_i = \sum_j S^T_{ij} \underline{\varphi}^\text{out}_j, \label{eq:scattdef2}
\end{align}
where the states  $\underline{\varphi}^\text{out}_i$ have to be constructed in analogy to $\underline{\varphi}^\text{in}_i$  in Sec.~\ref{sec:sol1}- \ref{sec:sol3}.
In case of time-reversal invariance, inscattering and outscattering states are time-reversal partners, $\underline{\varphi}^\text{out}_i=\left(\underline{\varphi}^\text{in}_i\right)^*$, and it immediately follows that 
\begin{align}
S^*S=\mathds{1}.\label{eq:Stimereversal}
\end{align}
Now, using this relation, together with $\underline{\varphi}^\text{out}_i=\left(\underline{\varphi}^\text{in}\right)^*_i$, as well as $\underline{m}_{\alpha,\pm}^*=\underline{m}_{\alpha,\mp}$ and $\bar{\underline{m}}_{\alpha,\pm}^*=\bar{\underline{m}}_{\alpha,\mp}$, it is straightforward to show that under time-reversal invariance the two definitions of $S$ from Eq.~\eqref{eq:scattdef1} and Eq.~\eqref{eq:scattdef2} are compatible.

Inserting Eq.~\eqref{eq:scattdef2} into  the magnon field operator expansion Eq.~\eqref{eq:bogansatz} allows us to identify the outscattering operators\footnote{To identify $a_i^\text{out}$, we used that  we can rewrite $\sum_i u^\text{in}_{\alpha,i} a^\text{in}_i= \sum_i u^\text{out}_{\alpha,i} a_i^\text{out}$ and  $\sum_i v^\text{in}_{\alpha,i} a^\text{in}_i= \sum_i v^\text{out}_{\alpha,i} a_i^\text{out}$ in Eq.~\eqref{eq:bogansatz}.}
\begin{align}
a_i^\text{out}=\sum_j S_{ij}a_j^\text{in}.\label{eq:scatteringmatrix}
\end{align}
Since $a_i^\text{out}$ and $a_i^\text{in}$ have identical commutation relations, the scattering matrix $S$ must fulfill
\begin{align}
    S\Sigma S^\dagger=\Sigma,\label{eq:Scommute}
\end{align}
where $\Sigma=\text{diag}(1,-1,1)$. 
Employing both Eq.~\eqref{eq:Stimereversal} and Eq.~\eqref{eq:Scommute} yields the relations $t_2=-\bar{t}_1$ and  ${t}^*_2 \bar{r}_2=\bar{t_1} r^*_1$ as well as 
\begin{align}
\abs{{r}_1}^2-\abs{ \bar{t}_1}^2=1\quad \text{and} \quad \abs{\bar{r}_2}^2-\abs{ {t}_2}^2=1.
\end{align}

\subsubsection{Two-mode squeezing}\label{sec:squeezing}
The relation  between $a^\text{in}_{1,2}$  and $a^\text{out}_{1,2}$ from Eq.~\eqref{eq:scatteringmatrix} given by the scattering matrix $S$ can also be formulated in Fock space using a two-mode squeezing transformation $U$ such that 
\begin{align}
e^{-i\phi_1} a^\text{out}_{1}=U^\dagger  a^\text{in}_{1} U= \cosh r \,a^\text{in}_{1}-e^{+i\theta} \sinh r \,  a^\text{in}_{2},\\
e^{-i\phi_2} a^\text{out}_{2}= U^\dagger  a^\text{in}_{2}U =\cosh r  \, a^\text{in}_{2}-e^{-i\theta} \sinh r   \,a^\text{in}_{1},
\end{align}
where we defined $\cosh(r)=\abs{r_1}$,  $\theta=\arg({t}^*_2 \bar{r}_2)$, and extracted the overall phases $\phi_1=\arg(r_1)$ and $\phi_2=\arg(\bar{r}_2)$. 
Note that the squeezing is hidden in this notation as the antimagnonic operators  $a^\text{in}_{2}$ and $a^\text{out}_{2}$ behave as creation operators. 
The unitary transformation is given by
\begin{align}
U=e^{\int\limits_0^\infty \mathrm{d}\varepsilon \left[ \xi_\text{in} ^*  a^\text{in}_{1}  \left(a^\text{in}_{2}\right)^\dagger -  \xi_\text{in}  \left(a^\text{in}_{1}\right)^\dagger  a^\text{in}_{2} \right]}=e^{\int\limits_0^\infty \mathrm{d}\varepsilon \left[ \xi_\text{out}^*  a^\text{out}_{1}  \left(a^\text{out}_{2}\right)^\dagger -  \xi_\text{out}  \left(a^\text{out}_{1}\right)^\dagger  a^\text{out}_{2} \right]},
\end{align}
where $\xi_\text{in}=r e^{i\theta}$ and $\xi_\text{out}=r e^{i\theta} e^{i(\phi_\text{L}-\phi_\text{R})}$. Then, our vacuum state of no incoming particles $\ket{\Omega}_\text{in}$ can be also expressed in terms of the outgoing basis via
\begin{align}
\ket{\Omega}_\text{in}=U \ket{\Omega}_\text{out} =\bigotimes\limits_{h_\text{L}<\varepsilon<h_\text{R}} \frac{1}{\abs{r_1}} \sum_n (-1)^n e^{i n (\theta+\phi_\text{L}-\phi_\text{R})} \left(\frac{\abs{\bar{t}_1}}{\abs{r_1}}\right)^n \ket{n,n,0}_\text{out},\label{eq:squeezed}
\end{align}
where $\ket{n_1,n_2,n_3}_\text{out}$ is the Fock state for $n_i$ magnons in the $i$-th outscattering state. The big circle denotes the continuous tensor product in the relevant energy window ${h_\text{L}<\varepsilon<h_\text{R}}$.  Beyond that window, the magnon occupation is zero, which we do not write explicitly. 
From Eq.~\eqref{eq:squeezed}, we identify  
\begin{align}
    p_n=\left(1-\frac{\vert \bar{t}_1\vert^2}{\vert r_1\vert^2}\right) \left(\frac{\vert \bar{t}_1\vert^2}{\vert r_1\vert^2}\right)^n,
\end{align}
which gives the probability of finding $n$ outgoing magnon-antimagnon pairs at energy $\varepsilon$. The probability follows a geometric probability distribution.

\section{Magnetic coupling to color centers}
Now, we also consider the color centers, which are described by the system Hamiltonian
\begin{align}
   H_\text{S} = - \frac{\boldsymbol{\Delta}_\text{L}}{\hbar} \cdot\frac{\hbar}{2}  \boldsymbol{\sigma_{\text{L}}}  - \frac{\boldsymbol{\Delta}_\text{R}}{\hbar} \cdot  \frac{\hbar}{2} \boldsymbol{\sigma_{\text{R}}} ,
\end{align}
where we assume $\boldsymbol{\Delta}_\text{L,R}=({0, 0, \Delta})$ with energy $\Delta>0$ which can be tuned through an external magnetic field. Thus, energetically both color centers want to point in the $+z$ direction. We denote the basis states of the color centers as $\ket{\text{g,g}} ,\ket{\text{g,e}} ,\ket{\text{e,g}}$ and $\ket{\text{e,e}}$ with $\text{g}$ denoting the ground state and $\text{e}$ the excited state. Without an environment, the ground state is $\ket{\text{g,g}}$ with both color centers pointing in the $z$-direction. However, the situation changes when we couple the color centers to the stray field of the magnetic environment via the coupling Hamiltonian 
\begin{align}
    H_\text{C}&= -\vb{h}_s(\vb{r}_\text{L})  \cdot  \frac{\hbar}{2} \boldsymbol{\sigma}_\text{L} -\vb{h}_s(\vb{r}_\text{R})  \cdot  \frac{\hbar}{2} \boldsymbol{\sigma}_\text{R}, \\
    &=- \frac{\hbar}{2} \sum\limits_{\alpha=\text{L,R}}  \left[ h^+_s(\vb{r}_\alpha)  \sigma^-_\alpha+h^-_s(\vb{r}_\alpha)  \sigma^+_\alpha  +h^z_s(\vb{r}_\alpha)  \sigma^z_\alpha  \right] , \label{eq:coupling}
\end{align}
where we assume that the color centers are positioned above the environment at distance $d$ on either side of the interface with $\vb{r}_\alpha=(x_\alpha,d,0)$, where $x_\text{L}<0$, $x_\text{R}>0$, and $d>0$. In the second line, we introduced $h_s^\pm= h^x_s\pm i h^y_s$ and $\sigma_\alpha^\pm=\left( \sigma^x_\alpha \pm i \sigma^y_\alpha\right)/2$.
\subsection{Stray magnetic field}
The stray field can be calculated via the magnetostatic Green's function $G_V(\vb{r}-\vb{r}^\prime)=-\nabla_{\vb{r}} \nabla_{\vb{r}^\prime} \frac{\gamma_e^2}{\abs{\vb{r}-\vb{r}^\prime}}$ ~\cite{guslienko_2011}
\begin{align}
 \vb{h}_s(\vb{r}_\alpha)=\int \mathrm{d}\vb{r}^\prime G_V(\vb{r}_\alpha-\vb{r}^\prime) \vb{s}_V(\vb{r}^\prime)=\int \mathrm{d}x^\prime G(x_\alpha-{x}^\prime) \frac{\vb{s}(x^\prime)}{\normfac}, \label{eq:green}
\end{align}
which fulfills the magnetostatic equation in matter $\nabla \cdot  \vb{h}_s= -4\pi \gamma_e^2 \nabla \cdot  \vb{s}_V$ (in Gaussian units; for SI units one has to replace $4\pi$ by the vacuum permeability $\mu_0$), where we again included the gyromagnetic ratio $\gamma_e$ in the definition of $\vb{h}_s$. The spin density per volume is given by 
\begin{align}
\vb{s}_V(\vb{r})= \frac{\Theta(\abs{z}-\frac{W}{2})}{\normfac} \delta(y) \vb{s}(x),
\end{align}
where $\vb{s}(x)=\left[ \Theta(-x) \vb{s}_\text{L}(x) +   \Theta(x) \vb{s}_\text{R}(x) \right]$ is the spin density per length derived in the scattering theory above. We normalize by $\normfac$ to make $\vb{s}_V(\vb{r})$ a proper volume density. In the second step of Eq.~\eqref{eq:green}, we introduced the one-dimensional Green's function 
\begin{align}
G(x_\alpha-{x}^\prime)= \int\limits_{-W/2}^{W/2} {\mathrm{d}z^\prime} G_V(x_\alpha-x^\prime,-d,-z^\prime). 
\end{align}
Using the identity $\abs{\vb{r}}^{-1}=\int \mathrm{d} k_x \mathrm{d} k_z e^{- k\abs{y}+ik_x x +i k_z z}/2\pi k$, we find 
\begin{align}
G({x}-x^\prime)&=-\frac{\gamma_e^2}{2\pi} \int \mathrm{d} k_x \int \mathrm{d} k_z \,   \frac{e^{-k d +i k_x (x-x^\prime) }}{k}W \text{sinc} \left( \frac{k_z W}{2}\right)
     \begin{pmatrix}
    k_x^2 &&    i k k_x &&  k_x k_z  \\
   i k k_x  &&  -k^2  && i k k_z \\
     k_x k_z  &&  i k k_z    &&k_z^2 
    \end{pmatrix} \\
    &=-{\gamma_e^2} \int \mathrm{d} k_x  {e^{-\abs{k_x} d +i k_x (x-x^\prime) }}
     \begin{pmatrix}
    \abs{k_x}  &&   i k_x  &&  0  \\
   i  k_x &&  -\abs{k_x}  && 0  \\
     0  && 0 &&0
    \end{pmatrix},
\end{align}
where $k=\sqrt{k_x^2+k_z^2}$. To get to the second line, we assume the magnetic field of a two-dimensional system $\lim_{W\rightarrow\infty} W \text{sinc} \left( \frac{k_z W}{2}\right)=2\pi \delta (k_z)$ motivated by the assumption $d\ll W$. 
Then, we obtain for the stray magnetic field $h^\pm_s=h^x_s\pm i h^y_s$ the expression
\begin{align}
h^\pm_s(\vb{r}_\alpha)= -\frac{2\gamma_e^2}{\normfac} \int\limits_0^\infty \mathrm{d} k k e^{-k d} \int\limits \mathrm{d} x   e^{\mp  i k (x_\alpha-x)} s^\mp(x),
\end{align}
where  $s^\pm_s=s^x_s\pm i s^y_s$. Thus, the overall stray field is diluted with $1/W$.
\begin{figure}[h]
\includegraphics[width=13cm]{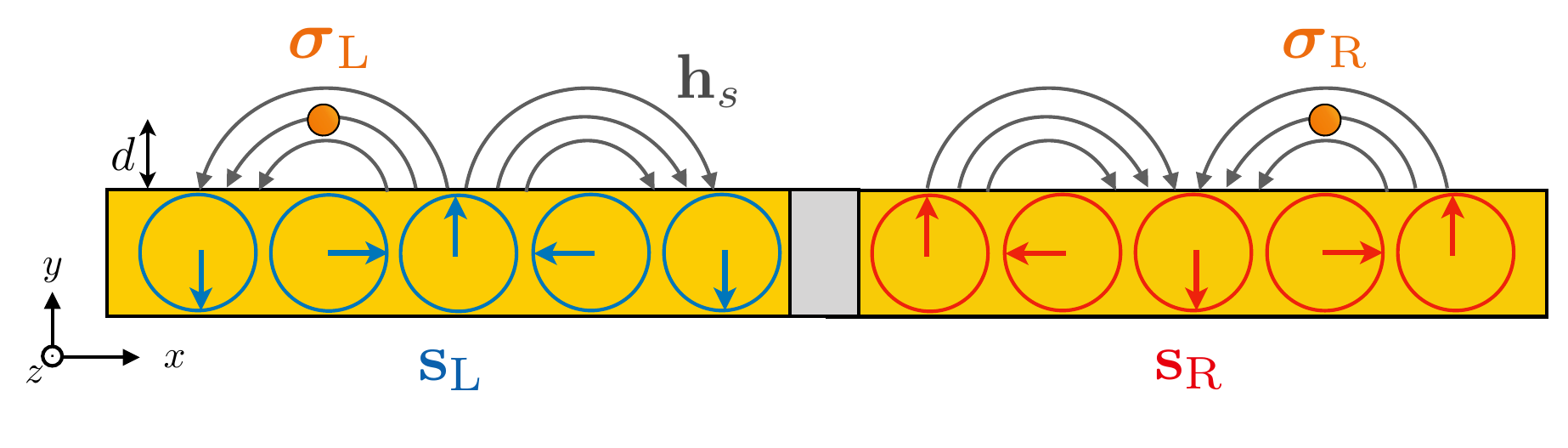}
	\caption{Snapshot of Halbach-like stray field. On the left, magnons with $k<0$ create a finite (zero) stray magnetic field above (below) the magnet. On the right, magnons with $k>0$ create a finite (zero) stray magnetic field above (below) the magnet. By flipping the sign of the magnon wave numbers on either side, the stray field will only be generated below the magnet (not shown). Thus, the coupling to the color centers (orange dots) is chiral.  For static magnetic field configurations, this effect is used in a Halbach array~\cite{halbach_1985}. }
	\label{fig:supp6}
\end{figure}
Interestingly, the coupling is of a chiral nature, see Fig.~\ref{fig:supp6}. 

In the following, we assume that $\abs{x_\alpha}\gg d$, such that the left (right) color center only probes the stray field generated by $\vb{s}_\text{L}$ ($\vb{s}_\text{R}$). Then, 
we obtain
\begin{align}
h^+_s(\vb{r}_\text{L})&\approx  -\frac{2\gamma_e^2}{\normfac} \sqrt{2\hbar s} \int \limits_{0}^{\infty}\frac{\mathrm{d}\varepsilon}{\sqrt{2\pi \hbar  v_\text{L}} } q_\text{L} e^{-q_\text{L} d}  e^{-  i q_\text{L} x_\text{L}}  \Bigg(r_1 a_{1} + t_2 a_{2}\Bigg), \\
h^+_s(\vb{r}_\text{R})&\approx  -\frac{2\gamma_e^2}{\normfac} \sqrt{2\hbar s} \int \limits_{0}^{\infty}\frac{\mathrm{d}\varepsilon}{\sqrt{2\pi \hbar  \bar{v}_\text{R}} }\left[ \bar{q}_\text{R} e^{-\bar{q}_\text{R} d}  e^{-  i \bar{q}_\text{R} x_\text{R}}  \Bigg(\bar{t}_1 a_{1} + \bar{r}_2 a_{2}\Bigg)  + {q}_\text{R} e^{-{q}_\text{R} d}  e^{-  i {q}_\text{R} x_\text{R}}\sqrt{\frac{\bar{v}_\text{R}}{{v}_\text{R}}} r^*_3  a_{3}^\dagger\right] \label{eq:strayfield},
\end{align}
where we neglected the evanescent modes on the left-hand side. Furthermore, also the last term of Eq.~\eqref{eq:strayfield} will not be important since it does not resonantly couple to the color centers. In particular, the coupling term $h^+_s(\vb{r}_\text{R}) \sigma^-_\text{R}$ from Eq.~\eqref{eq:coupling} gives rise to a strongly oscillating term $a_3^\dagger(\varepsilon) \sigma_- \sim e^{i(\varepsilon+\Delta) t/\hbar}$ with $\varepsilon>0$ and $\Delta>0$. Thus, the third scattering state $a_3$ is irrelevant for the coupling in the secular approximation.

\subsection{Master equation}
Integrating out the environment's degrees of freedom, we obtain the master equation
\begin{align}
\partial_t \rho = \frac{1}{i\hbar} \left[ H_\text{S}+\delta H,\rho\right] + \sum_{i,j=1}^4 \Gamma_{ij}\left( E_i \rho E_j^\dagger -\frac{1}{2} \{ E^\dagger_j E_i, \rho \} \right),
\end{align}
where $\delta H$ describes the environment-induced coherent dynamics and $\Gamma$ the dissipative dynamics. As an operator basis we have chosen $\left( E_1,E_2,E_3,E_4\right)=\left( \sigma^+_\text{L} , \sigma^+_\text{R},\sigma^-_\text{L} , \sigma^-_\text{R}\right)$. Note that we use the convention that $\sigma^+_\alpha$ always increases the spin in the $z$-direction and, thereby, \textit{lowers} the energy.
\subsubsection{Dissipative coupling}
For the dissipative coupling,  we obtain  
\begin{align}
\Gamma=\Gamma_1 \oplus \Gamma_2 =\frac{1}{4}
\begin{pmatrix}
   i\,  G^>_\text{LL} && i \,G^>_\text{LR} && 0  &&0\\
  i  \, G^>_\text{RL} &&  i \,G^>_\text{RR} && 0&& 0 \\
     0 && 0 &&  i \,G^<_\text{LL}   &&i \,G^<_\text{RL}\\
     0 && 0 && i \,G^<_\text{LR}&&i \, G^<_\text{RR}
    \end{pmatrix},
\end{align}
where the $1/4$ originates from the coupling Hamiltonian from Eq.~\eqref{eq:coupling}. The $\hbar$ cancels in the derivation. 
Then, the decay rates are given by the correlators of the stray magnetic field $\vb{h}_s$ in the form of lesser and larger Green's functions~\cite{zou_2022}
\begin{align}
G^>_{\alpha \beta} (\Delta)= -i \int \mathrm{d} \tau e^{i \Delta \tau/\hbar}  \mel{\Omega}{h_s^+(\vb{r}_\alpha,\tau) h_s^-(\vb{r}_\beta)} {\Omega},\\
G^<_{\beta \alpha} (\Delta)= -i \int \mathrm{d} \tau e^{i \Delta \tau/\hbar}  \mel{\Omega}{h_s^-(\vb{r}_\alpha)h_s^+(\vb{r}_\beta,\tau) } {\Omega}. 
\end{align}
Note that the full matrix decouples into two $2\times2$  blocks $\Gamma_1$ and $\Gamma_2$ which only originate from processes involving the scattering states $a_1$ and $a_2$, respectively. 
In our case, these blocks can be written in the simple form $\Gamma_{1}=\Gamma_0 \boldsymbol{\lambda}_1\boldsymbol{\lambda}_1^\dagger$ and $\Gamma_{2}=\Gamma_0 \boldsymbol{\lambda}_2 \boldsymbol{\lambda}_2^\dagger$ with
\begin{align}
\boldsymbol{\lambda}_1=\begin{pmatrix}    \sqrt{q_\text{L}d}\, e^{-q_\text{L} d}e^{-i q_\text{L} x_\text{L}} r_1 \\  \sqrt{\bar{q}_\text{R}d}\,e^{-\bar{q}_\text{R} d} e^{-i \bar{q}_\text{R} x_\text{R}} \bar{t}_1 \end{pmatrix}, \quad \boldsymbol{\lambda}_2=\begin{pmatrix}    \sqrt{q_\text{L}d} \,e^{-q_\text{L} d} e^{i q_\text{L} x_\text{L}}{t}^*_2 \\ \sqrt{\bar{q}_\text{R}d}\,e^{-\bar{q}_\text{R} d}  e^{i \bar{q}_\text{R} x_\text{R}}\bar{r}^*_2 \end{pmatrix},\label{eq:complexvector}
\end{align}
where we defined the constant $\Gamma_0=\frac{\gamma_e^4 \hbar^2 s}{\normfac^2 Ad}$ (in Gaussian units). In Eq.\eqref{eq:complexvector}, all functions are evaluated at energy $\Delta$. 
As a next step, we eliminate the phases by performing a gauge transformation (rotation around the $z$-axis) of the form
\begin{align}
\tilde{\sigma}_\text{L}^\pm=e^{\pm i \left(\text{arg} \,r_1- q_\text{L} x_\text{L} \right)} \sigma_\text{L}^\pm,\\
\tilde{\sigma}_\text{R}^\pm=e^{\pm i \left( \text{arg} \,\bar{t}_1-\bar{q}_\text{R} x_\text{R} \right)} \sigma_\text{R}^\pm.
\end{align}
By using that $r_1 {t}_2^*=\bar{r}_2^*\bar{t}_1$, which follows from the properties of the scattering matrix shown in Eq.\eqref{eq:Stimereversal}-\eqref{eq:Scommute}, we get the real matrices
\begin{align}
\Gamma_1=\Gamma_0 d 
\begin{pmatrix}
 q_\text{L} e^{-2q_\text{L} d} \,r^2 && \sqrt{ q_\text{L}  \bar{q}_\text{R}} e^{-(q_\text{L} +\bar{q}_\text{R})d} \, r  t \\
 \sqrt{ q_\text{L}  \bar{q}_\text{R}} e^{-(q_\text{L} +\bar{q}_\text{R})d}\, r  t    &&   \bar{q}_\text{R} e^{-2 \bar{q}_\text{R}  d}\,t^2 
 \end{pmatrix},\\
 \Gamma_2=\Gamma_0 d 
 \begin{pmatrix}
    q_\text{L} e^{-2q_\text{L} d} \,t^2 && \sqrt{ q_\text{L}  \bar{q}_\text{R}} e^{-(q_\text{L} +\bar{q}_\text{R})d} \, r  t     \\
      \sqrt{ q_\text{L}  \bar{q}_\text{R}} e^{-(q_\text{L} +\bar{q}_\text{R})d}\, r  t    &&   \bar{q}_\text{R} e^{-2 \bar{q}_\text{R}  d}\,r^2 
    \end{pmatrix},
 \end{align}
 where $  r =\abs{r_1}=\abs{\bar{r}_2}$ and $t=\abs{\bar{t}_1}=\abs{{t}_2}$. Note that the gauge transformation has fully removed the dependence on the position $x_\alpha$ of the color centers. 
 
Since the matrices $\Gamma_{1}$ and $\Gamma_{2}$ are formed by outer products of the vectors $\boldsymbol{\lambda}_1$  and $\boldsymbol{\lambda}_2$, one eigenvalue is always zero. Thus, the diagonalized Lindblad equation has only two relevant Lindblad  operators
\begin{align}
\partial_t \rho = \frac{1}{i\hbar} \left[ H_\text{S}+\delta H,\rho\right] + \sum_{i=1}^2 \Gamma_0 \left( L_i \rho L_i^\dagger -\frac{1}{2} \{ L^\dagger_i L_i, \rho \} \right),
\end{align}
 with
 \begin{align}
 L_1&=  \sqrt{q_\text{L}d}\, e^{-q_\text{L} d} r\, \tilde{\sigma}^+_\text{L}+ \sqrt{\bar{q}_\text{R}d}\,e^{-\bar{q}_\text{R} d}  t \,\tilde{\sigma}^+_\text{R} \\
  L_2&=  \sqrt{q_\text{L}d}\, e^{-q_\text{L} d} t\,\tilde{\sigma}^-_\text{L}+ \sqrt{\bar{q}_\text{R}d}\,e^{-\bar{q}_\text{R} d}  r \,\tilde{\sigma}^-_\text{R}.
  \end{align}
We can see already a remarkable effect even in the case where the two magnets of the environment are not coupled  ($r=1$ and $t=0$). While the left color center does decay to its ground state with $L_1\propto \sigma_\text{L}^+$, the right color center will always transition into its excited state with $L_2 \propto \sigma_\text{R}^-$. Thus, the steady state will be $\ket{\text{g,e}}$ instead of $\ket{\text{g,g}}$. This reflects the fact that the right color center can always emit a negative-energy magnon into the inverted magnet and hence completely reverse its occupation. 
\subsubsection{Coherent coupling}
Besides the dissipative coupling, the environment generically also mediates a nonlocal coherent coupling similar to RKKY interaction of the form
\begin{align}
\delta H= J \, \sigma^+_\text{L}  \sigma^-_\text{R} +   J^*\, \sigma^+_\text{R}  \sigma^-_\text{L} ,
\end{align}
where
\begin{align}
J= \frac{1}{4} \mathfrak{Re}\, G^\text{R}_{\text{RL}}(\Delta).
\end{align}
Here we followed Ref.~\cite{zou_2022} and defined a ``real" part of the retarded Green's function which is given by
\begin{align}
\mathfrak{Re}\,G^\text{R}_{\text{RL}}(\Delta)=-i \int \mathrm{d} \tau \, \text{sgn}(\tau) \, e^{i\Delta \tau/\hbar}  \mel{\Omega}{\left[ h_s^+(\vb{r}_\text{R},\tau) ,h_s^-(\vb{r}_\text{L})\right]}{\Omega} =0,
\end{align}
where $\text{sgn}(\tau)$ is the signum function. For our system, we have $\left[ h_s^+(\vb{r}_\text{R},\tau) ,h_s^-(\vb{r}_\text{L})\right]=0$, where we made use of the relation $r_1 \bar{t}^*_1=t_2 \bar{r}_2^*$ that follows from Eq.~\eqref{eq:Scommute} for the scattering matrix. Thus, there is no nonlocal coherent coupling between the color centers. 
There is, however, a local Lamb shift  $\propto \delta_\alpha \sigma_\alpha^z$ on either side, which renormalizes the bare energies $\Delta$. This contribution will be ignored in this work as only differences $\delta_\text{L}-\delta_\text{R}$ will change our results.

\subsection{Entanglement}
For the energy $\Delta=(h_\text{L}+h_\text{R})/2$, i.e. in the middle of the magnon-antimagnon emission spectrum, we obtain $q_\text{L}=\bar{q}_\text{R}=q$. Then, the Lindblad operators take on the simple form
 \begin{align}
 L_1&=  \sqrt{q d}\, e^{-q d} \left(r\, \tilde{\sigma}^+_\text{L}+  t \,\tilde{\sigma}^+_\text{R} \right),\\
  L_2&=  \sqrt{q d}\, e^{-q d} \left(t\,\tilde{\sigma}^-_\text{L}+   r \,\tilde{\sigma}^-_\text{R} \right).
  \end{align}
  The steady state is pure and has a particularly simple form given by
\begin{align}
\rho_\text{st}=\dyad{\psi},\quad\text{with} \quad \ket{\psi}=\frac{r\ket{\text{g,e}}-t\ket{\text{e,g}}}{\sqrt{r^2+t^2}}.
\end{align}
The pair creation occurs for $t>0$ and leads to the admixture of the state $\ket{\text{e,g}}$, enabling entanglement between the color centers.

\section{Exchange coupling to color centers}
As another example, we consider color centers coupled to the magnets via exchange interaction:
\begin{align}
   H_\text{S} &= - \frac{\boldsymbol{\Delta}_\text{L}}{\hbar} \cdot\frac{\hbar}{2}  \boldsymbol{\sigma_{\text{L}}}  - \frac{\boldsymbol{\Delta}_\text{R}}{\hbar} \cdot  \frac{\hbar}{2} \boldsymbol{\sigma_{\text{R}}}  \\ 
       H_\text{C}&= - J_X \left[ \vb{s}_\text{L}(x_\text{L}) \cdot  \frac{\hbar}{2} \boldsymbol{\sigma}_\text{L} +\vb{s}_\text{R}(x_\text{R})  \cdot  \frac{\hbar}{2} \boldsymbol{\sigma}_\text{R}\right] \\
    &=- \frac{\hbar  J_X  }{2} \sum\limits_{\alpha=\text{L,R}}  \left[ s^+_\alpha(x_\alpha)  \sigma^-_\alpha+s^-_\alpha(x_\alpha)  \sigma^+_\alpha  +s^z_\alpha(x_\alpha)  \sigma^z_\alpha  \right].
\end{align}
Here, we assume flipped color centers with $\boldsymbol{\Delta}_\text{L,R}=({0, 0, -\Delta})$ and $\Delta>0$ such that energetically both color centers want to point in the $-z$ direction. Then, $\sigma_\alpha^+$ and $\sigma_\alpha^-$ increase and decrease the energy, respectively. The second line describes the local exchange interaction parametrized by the exchange coupling $J_X$. Since the longitudinal fluctuations in $s^z_\alpha$ are of higher order, they are going to be ignored in the following. 
Carrying out an analogous calculation as in the last section, we find the Lindblad operators
 \begin{align}
 L_1&=  \frac{1}{\sqrt{q_\text{L}\ell}}\,(e^{i q_\text{L} x_\text{L}}+r_1e^{-i q_\text{L} x_\text{L}})\,{\sigma}^-_\text{L}+  \frac{1}{\sqrt{\bar{q}_\text{R}\ell}} \, \left( \bar{t}_1e^{-i \bar{q}_\text{R}  x_\text{R}}\right) \,{\sigma}^-_\text{R} \\
  L_2&=  \frac{1}{\sqrt{{q}_\text{L}\ell}}\,\left(t_2 e^{-i q_\text{L} x_\text{L}}\right)^*\,{\sigma}^+_\text{L}+\frac{1}{\sqrt{\bar{q}_\text{R}\ell}}(e^{i \bar{q}_\text{R}  x_\text{R}}+ \bar{r}_2e^{-i \bar{q}_\text{R}  x_\text{R}})^* \,{\sigma}^+_\text{R},
  \end{align}
  where the rate is now given by $\Gamma_0=J_X^2 \hbar^2 s \ell /4A$. 
  
\begin{figure}[h]
\includegraphics[width=13cm]{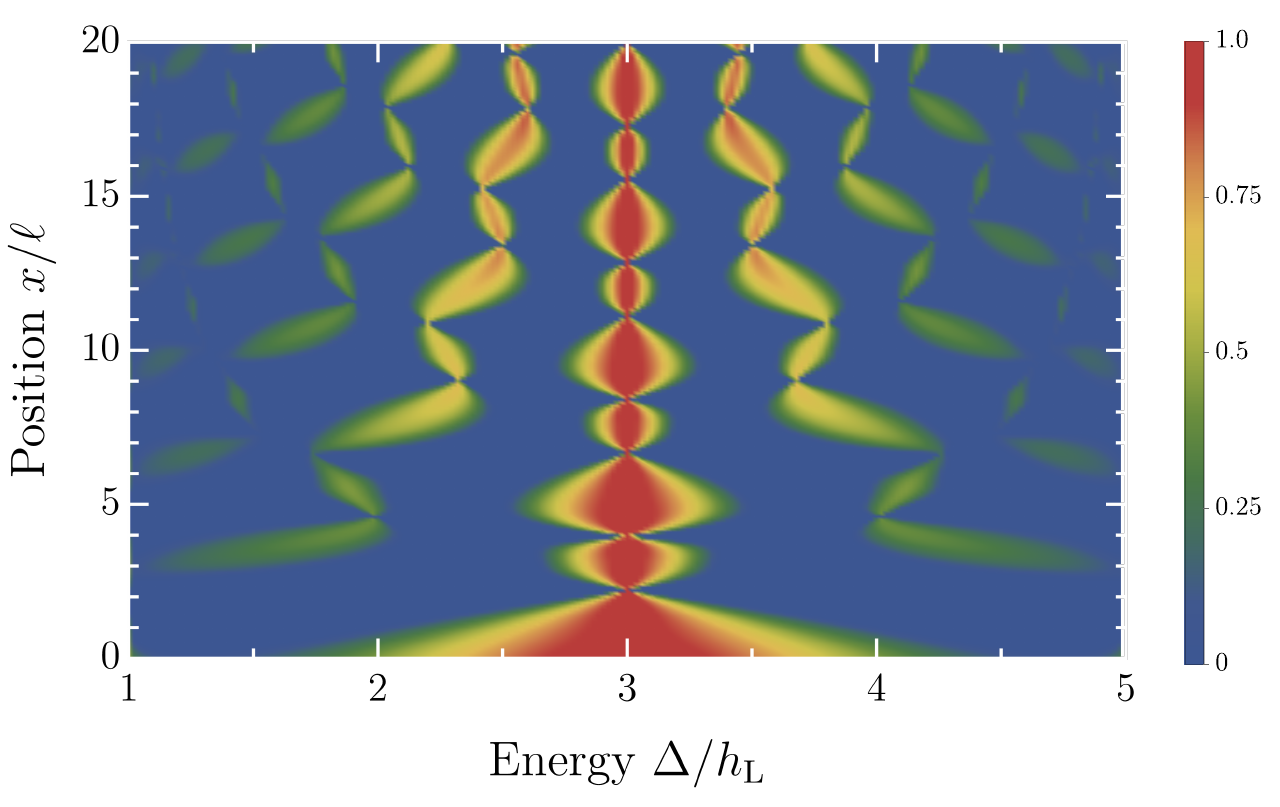}
	\caption{Steady-state concurrence ${\cal C}$ (an entanglement measure) as a function of the color center energy $\Delta$ and position $x=-x_\text{L}=x_\text{R}$. In contrast to the chiral magnetostatic coupling in the Letter, the exchange coupling is nonchiral, and hence a dependence on the color center positions $x_\text{L}$ and $x_\text{R}$ survives. We choose $h_\text{R}=5 h_\text{L}$, $\gamma=A/\ell$ and define the length scale $\ell=\sqrt{\stiff/(h_\text{R}{-}h_\text{L})}$.}
	\label{fig:supp7}
\end{figure}

   In Fig.~\ref{fig:supp7}, we show the entanglement of the steady state resulting from these Lindblad operators (where we ignore environment-induced coherent effects from $\delta H$).
   As a measure for entanglement in a given stationary state $\rho_\text{st}$, we use the concurrence ${\cal C}(\rho_\text{st})$ defined for two-qubits systems by~\cite{wooters_1998} 
   \begin{align}
   {\cal C}(\rho_\text{st})=2\text{max}\left(0,\abs{\mel{\text{e,g}}{\rho_\text{st}}{\text{g,e} }-\sqrt{\mel{\text{g,g}}{\rho_\text{st}}{\text{g,g} }\mel{\text{e,e}}{\rho_\text{st}}{\text{e,e} }}}\right).
   \end{align}
   Here, we used that for $\Delta\neq 0$ the only surviving coherence in the stationary state is between $\ket{\text{g,e}}$ and $\ket{\text{e,g}}$.
    
    Since the exchange coupling is nonchiral, the spatial dependence on the color centers   $x_\text{L}$ and $x_\text{R}$ survives and entanglement occurs at certain resonances determined by the energy $\Delta$ and the positions $x_\text{L}$ and $x_\text{R}$. Thus, the chiral nature of the magnetostatic coupling is helpful, but not essential to imprint entanglement onto the color centers. 

\bibliography{References}
